\documentclass[10pt,twocolumn,english,aps,prb]{revtex4-1}
\usepackage[latin9]{inputenc}
\usepackage{amsmath}
\usepackage{amssymb}
\usepackage{graphicx}
\usepackage{esint}

\makeatletter

\@ifundefined{textcolor}{}
{%
 \definecolor{BLACK}{gray}{0}
 \definecolor{WHITE}{gray}{1}
 \definecolor{RED}{rgb}{1,0,0}
 \definecolor{GREEN}{rgb}{0,1,0}
 \definecolor{BLUE}{rgb}{0,0,1}
 \definecolor{CYAN}{cmyk}{1,0,0,0}
 \definecolor{MAGENTA}{cmyk}{0,1,0,0}
 \definecolor{YELLOW}{cmyk}{0,0,1,0}
}

\usepackage{mathrsfs}
\usepackage{bbm}
\usepackage{bbold}
\usepackage{subfigure}
\usepackage{epstopdf}
\usepackage{xcolor}
\usepackage{babel}

\newcommand{\rmd}{{\rm d}}

\newcommand{\fvec}[1]{\boldsymbol{#1}}

\newcommand{\dif}{\mathrm{d}}

\newcommand{\half}{\frac{1}{2}}
\newcommand{\pd}{\partial}

\usepackage{babel}

\usepackage{babel}

\makeatother

\usepackage{babel}
\begin{document}

\title{Robustness of quantum critical pairing against disorder}

\author{Jian Kang}

\email{jkang@umn.edu}

\selectlanguage{english}%

\affiliation{School of Physics and Astronomy, University of Minnesota, Minneapolis,
MN 55455, USA}

\author{Rafael M. Fernandes}

\affiliation{School of Physics and Astronomy, University of Minnesota, Minneapolis,
MN 55455, USA}
\begin{abstract}
The remarkable robustness of high-temperature superconductors against
disorder remains a controversial obstacle towards the elucidation
of their pairing state. Indeed, experiments report a weak suppression
rate of the transition temperature $T_{c}$ with disorder, significantly
smaller than the universal value predicted by extensions of the conventional
theory of dirty superconductors. However, in many high-$T_{c}$ compounds,
superconductivity appears near a putative magnetic quantum critical
point, suggesting that quantum fluctuations, which suppress coherent
electronic spectral weight, may also promote unconventional pairing.
Here we investigate theoretically the impact of disorder on such a
quantum critical pairing state, considering the coupling of impurities
both to the low-energy electronic states and to the pairing interaction
itself. We find a significant reduction in the suppression rate of
$T_{c}$ with disorder near the magnetic quantum critical point, shedding
new light on the nature of unconventional superconductivity in correlated
materials.
\end{abstract}
\maketitle

\section{Introduction}

\begin{figure}[htbp]
\includegraphics[width=0.95\columnwidth]{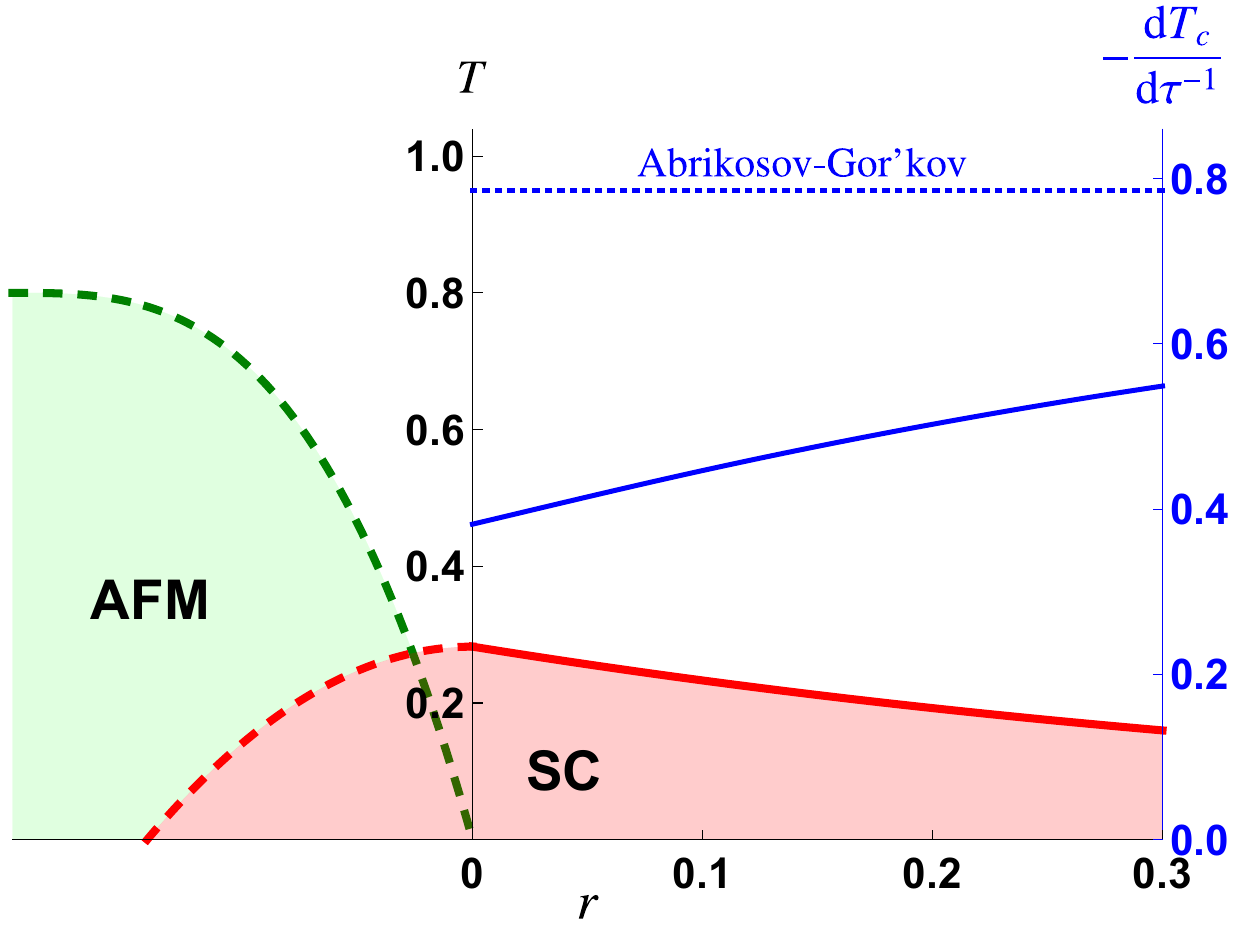} \protect\protect\caption{Phase diagram displaying a SC dome near an AFM-QCP. The dashed lines
are schematic, whereas the solid lines are the results of our calculations.
The red solid line denotes the SC transition temperature of the clean
system, $T_{c}$, as function of the distance to the QCP, $r$. Both
quantities are in units of the paring energy scale $\Lambda$ for
a cutoff $\Omega_{c}=3\Lambda$ (see text). The blue solid line denotes
the suppression rate of $T_{c}$ with pair-breaking scattering $\tau^{-1}=\tau_{Q}^{-1}$,
$\rmd T_{c}/\rmd\tau^{-1}$, due to the coupling between disorder
and the low-energy electronic states. The blue dotted line is the
standard AG universal value $\left(\rmd T_{c}/\rmd\tau^{-1}\right)_{\mathrm{AG}}=-\pi/4$.}

\label{Fig:QCPSC}
\end{figure}

Elucidating the nature of unconventional superconductivity (SC) remains
a major challenge in condensed matter physics. The fact that unconventional
SC is found in proximity to a magnetic instability in many heavy-fermion~\cite{HFSC,UPT3},
organic~\cite{OGSC,Soto95}, cuprate~\cite{CuSC}, and iron-based
compounds~\cite{FeSC}, led to the proposal that magnetic fluctuations
promote the binding of the electrons in Cooper pairs, resulting in
unconventional gap functions that change sign across the Brillouin
zone (such as $d$-wave and $s^{+-}$-wave gaps)~\cite{Hirsch85,Varma86,Millis88,Scalapino86,Leggett93,PhaseCuSC,Ketchen94,Pines92,Pines07,Mazin08,Kuroki08,Hirschfeld11,Scalapino12,AVC12}.
Indeed, in the phase diagram of high-temperature superconductors such
as electron-doped cuprates and iron pnictides, the maximum value of
$T_{c}$ is observed very close to a putative antiferromagnetic (AFM)
quantum critical point (QCP) \cite{Sachdev_book,Koshizuka99,Xu09,FeSCQCP},
as shown in Fig.~\ref{Fig:QCPSC}. Consequently, the possibility
of pairing mediated by quantum critical fluctuations has been extensively
investigated recently~\cite{AVC99,AVC03,Sachdev10,Zaanen11,Berg12,Pepin13,AVC13,Senthil15,Raghu15,Kivelson15}.

Experimentally, a major tool to probe unconventional SC has been the
behavior of $T_{c}$ with disorder \cite{Balatsky06}. In conventional
superconductors displaying a $s$-wave gap, weak non-magnetic impurity
scattering is known to be inconsequential to $T_{c}$ \cite{Anderson},
whereas magnetic impurities suppress $T_{c}$ according to the Abrikosov-Gor'kov
(AG) expression \cite{AG}. For a small pair-breaking scattering rate
$\tau^{-1}$, AG yields the universal suppression rate $\left(\rmd T_{c}/\rmd\tau^{-1}\right)_{\mathrm{AG}}=-\pi/4$,
confirmed experimentally \cite{WoolfImpExp} (see Fig.~\ref{Fig:QCPSC}).
Qualitatively, extensions of the AG theory to $d$-wave and $s^{+-}$
superconductors reveal that non-magnetic impurities are in general
pair-breaking. However, quantitatively, the experimentally observed
suppression of $T_{c}$ with disorder in cuprates and pnictides is
rather small compared to the AG-based results~\cite{Phillips96,Rullier03,Uchida05,Hirschfeld09,Murakami10,Paglione12}.
Several scenarios have been proposed to reconcile this robustness
of SC against disorder, including strong correlation effects \cite{Trivedi08,Miranda15},
spatial inhomogeneity of the gap function \cite{Franz97}, spin-fluctuation
mediated pairing \cite{AVC10,Norman93}, disorder-induced enhancement
of magnetic fluctuations \cite{Cheng09}, distinct intra-orbital and
inter-orbital scattering \cite{Zhu13,Mishra13,Kontani13,Jorg15},
and even models advocating for a standard $s$-wave gap in the pnictides
\cite{Kontani09}.

In this paper, to shed light on our understanding of unconventional
SC, we focus on how disorder affects critical AFM-mediated pairing
beyond the AG paradigm. In particular, we consider the impact of disorder
on a general spin-fermion model that describes SC promoted by quantum-critical
AFM fluctuations, which can be applied to both cuprates and pnictides.
Previously, Ref.~\cite{AVC10} found a strikingly resilient SC state
against the effects of disorder near an AFM-QCP. Here, instead of
solving the coupled Eliashberg equations \cite{Eliashberg}, we express
them in a convenient functional form \cite{BRMethod,Valls78,Millis88,RMF13}
that allows us to compute directly $\rmd T_{c}/\rmd\tau^{-1}$ and
gain invaluable insight on the different processes by which weak impurity
scattering affects $T_{c}$. Specifically, in the limit of weak scattering,
three independent contributions arise:

\begin{equation}
\frac{\rmd T_{c}}{\rmd\tau^{-1}}=\left(\frac{\rmd T_{c}}{\rmd\tau^{-1}}\right)_{f}+\left(\frac{\rmd T_{c}}{\rmd\tau^{-1}}\right)_{b,1}+\left(\frac{\rmd T_{c}}{\rmd\tau^{-1}}\right)_{b,2}\label{main}
\end{equation}

The first term, which yields the results of Fig.~\ref{Fig:QCPSC},
arises from the direct coupling of disorder and the low-energy electronic
states. This coupling leads to a decrease of the electronic coherent
spectral weight near the QCP, which in turn suppresses the reduction
of the pairing vertex caused by pair-breaking scattering, in agreement
with the general results from Ref.~\cite{AVC10}. In particular,
at the QCP, we find the value $\left(\frac{\rmd T_{c}}{\rmd\tau_{Q}^{-1}}\right)_{f}\approx-0.45$,
which is about half of the value expected from AG theory. The last
two terms in the equation above arise from the coupling of disorder
and the bosonic degrees of freedom that promote the pairing interaction
-- in this case, spin fluctuations. In general, $\left(\frac{\rmd T_{c}}{\rmd\tau^{-1}}\right)_{b,2}<0$
comes from the suppression of the correlation length of the quantum
critical fluctuations by disorder. On the other hand, $\left(\frac{\rmd T_{c}}{\rmd\tau^{-1}}\right)_{b,1}>0$
appears due to the renormalization of the electron-boson vertex, and
is generally larger than $\left|\left(\frac{\rmd T_{c}}{\rmd\tau^{-1}}\right)_{b,2}\right|$.
Consequently, the impact of disorder on the pairing interaction leads
to an additional reduction of $\rmd T_{c}/\rmd\tau^{-1}$ with respect
to the AG value. Our results offer a fresh perspective on the robustness
of unconventional SC against disorder, lending support to the proposal
that quantum critical pairing plays an important role in copper- and
iron-based SC.

The paper is organized as follows: Section II introduces the spin-fermion
model and the SC gap equations. Section III discusses the coupling
between disorder and the fermionic degrees of freedom, which yields
$\left(\frac{\rmd T_{c}}{\rmd\tau^{-1}}\right)_{f}$, whereas Section
IV discusses the coupling between disorder and the bosonic degrees
of freedom, which yields $\left(\frac{\rmd T_{c}}{\rmd\tau^{-1}}\right)_{b_{1}}$
and $\left(\frac{\rmd T_{c}}{\rmd\tau^{-1}}\right)_{b_{2}}$. Section
V is devoted to the concluding remarks. Analytical approximations
to $\left(\frac{\rmd T_{c}}{\rmd\tau^{-1}}\right)_{b_{1}}$ and $\left(\frac{\rmd T_{c}}{\rmd\tau^{-1}}\right)_{b_{2}}$
are given in Appendix A.

\section{Spin-fermion model and the linearized gap equations}

\begin{figure}[htbp]
\subfigure[\label{Fig:FS:Cu}]{\includegraphics[width=0.45\columnwidth]{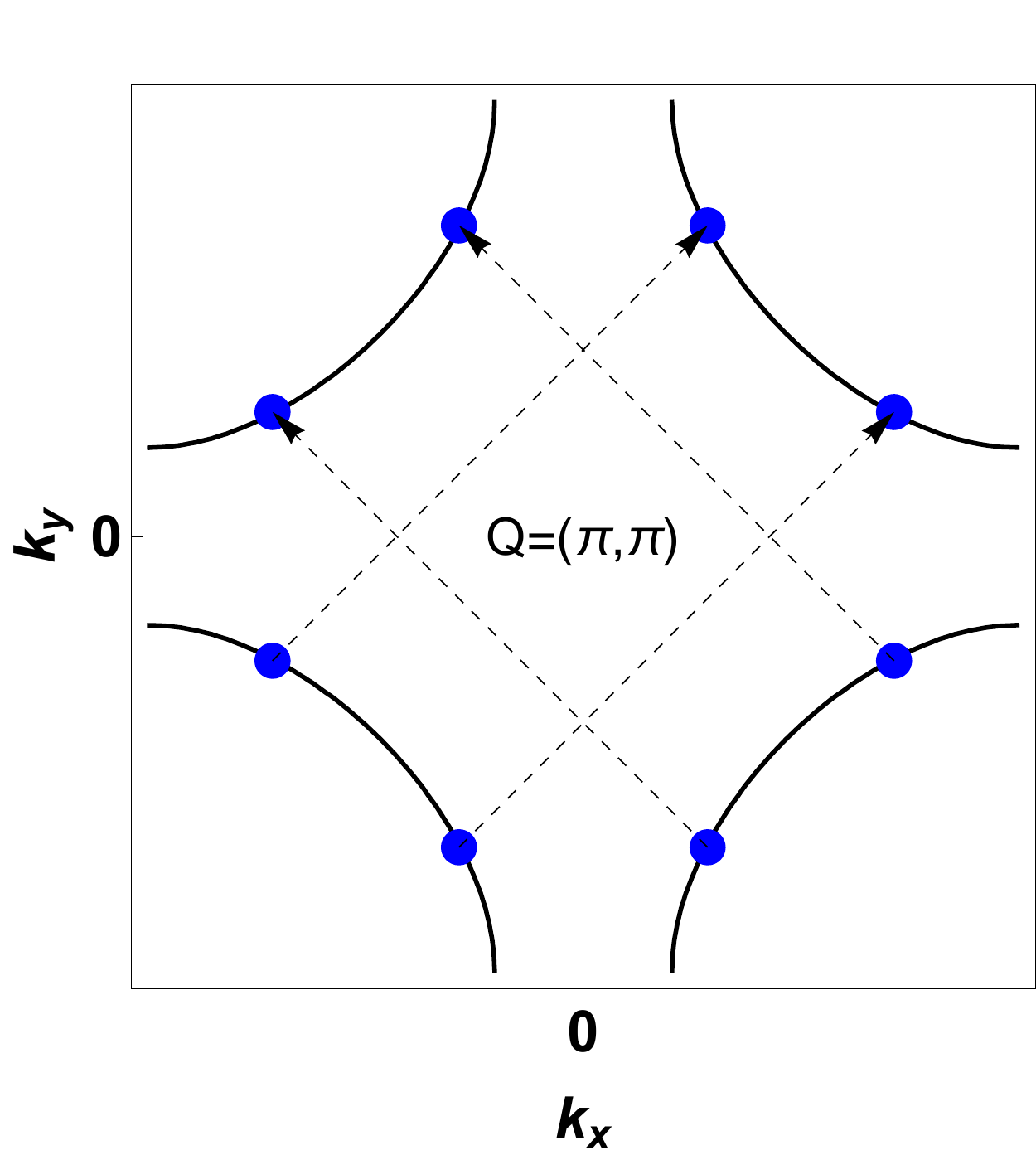}}
\hfill{} \subfigure[\label{Fig:FS:Iron}]{\includegraphics[width=0.45\columnwidth]{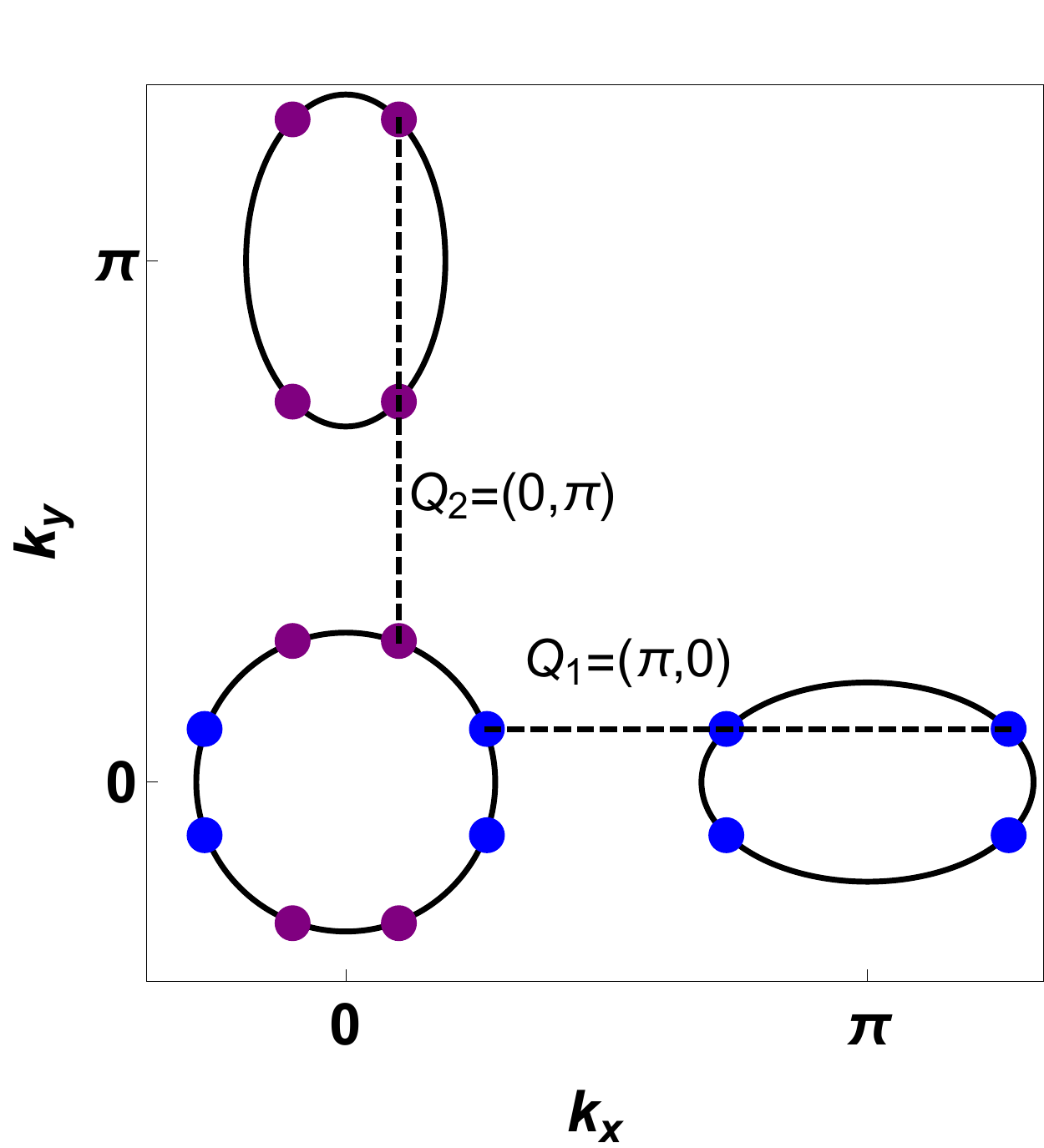}}
\protect\protect\caption{Schematic Fermi surfaces of (a) the cuprates and (b) iron pnictides,
respectively. Pairs of hot spots (blue or purple points) are connected
by dashed lines, corresponding to the momentum $\fvec Q=(\pi,\pi)$,
for the cuprates, and $\fvec Q=(\pi,0)$ or $(0,\pi)$, for the pnictides.
Spin fluctuations are peaked at these wave-vectors in the two materials.}

\label{Fig:FS}
\end{figure}

Our starting point is the low-energy spin-fermion model, in which
electrons couple to a bosonic AFM order parameter $\fvec\phi_{q}$,
whose fluctuations are described by the magnetic susceptibility $\chi_{b}(\fvec q,\Omega_{n})$.
We focus on the electronic states $c_{\fvec k\sigma}$ and $d_{\fvec k\sigma}\equiv c_{\fvec k+\fvec Q\sigma}$
in the vicinities of a pair of hot spots, i.e. points of the Fermi
surface connected by the AFM ordering vector $\fvec Q$. The action
is given by \cite{AVC03,AVC10}:
\begin{eqnarray}
 &  & S=\int_{k}(-i\omega_{n}+\epsilon_{c}(\fvec k))c_{\fvec k\sigma}^{\dag}c_{\fvec k\sigma}+\int_{k}(-i\omega_{n}+\epsilon_{d}(\fvec k))d_{\fvec k\sigma}^{\dag}d_{\fvec k\sigma}+\nonumber \\
 &  & \lambda\int_{k,q}\fvec\phi_{-q}\cdot\left(c_{k,\alpha}^{\dag}\fvec\sigma_{\alpha\beta}d_{\fvec k+\fvec q,\beta}\right)+\int_{q}\chi_{b}^{-1}(\fvec q,\Omega_{n})\fvec\phi_{q}\cdot\fvec\phi_{-q}\label{spin_fermion}
\end{eqnarray}
where $\int_{k}=T\sum_{n}\int\frac{\rmd^{d}k}{(2\pi)d}$, $\lambda$
is the coupling constant, $\epsilon_{d}(\fvec k)\equiv\epsilon_{c}(\fvec k+\fvec Q)$,
and $\omega_{n}=(2n+1)\pi T$ is the fermionic Matsubara frequency.
Because the behavior of this action is dominated by the states near
the hot spots \cite{Berg12}, we linearize the spectrum near them,
$\epsilon_{c}(\fvec k)\approx\fvec v_{c}\cdot\fvec k$ and $\epsilon_{d}(\fvec k)\approx\fvec v_{d}\cdot\fvec k$,
where $\fvec k$ is measured with respect to the Fermi momentum. Thus,
by focusing on a single pair of hot spots, this model can in principle
be applied to either cuprates or pnictides. Indeed, as shown in Fig.~\ref{Fig:FS},
there are four pairs of hot spots in the typical Fermi surface of
the cuprates (in which $\fvec Q=(\pi,\pi)$) and eight for the iron
pnictides (in which $\fvec Q=(\pi,0)$ or $(0,\pi)$). Hereafter,
for simplicity, we consider the special case $|\fvec v_{c}|=|\fvec v_{d}|$,
but the main results should remain valid otherwise.

For such a low-energy model, the magnetic susceptibility can be expanded
as $\chi_{b}^{-1}(\fvec q,\Omega_{n})=\chi_{0}^{-1}\left(r_{0}+q^{2}+\Omega_{n}^{2}/v_{b}^{2}\right)$,
where $\chi_{0}^{-1}$ is the magnetic energy scale determined by
high-energy states, $r_{0}$ is the distance to the bare AFM quantum
critical point, and $v_{b}$ is the spin-wave velocity. The coupling
to the electronic degrees of freedom, however, fundamentally changes
this propagator by introducing Landau damping, i.e. the decay of magnetic
excitations in electron-hole pairs. Within one-loop, the renormalized
magnetic susceptibility becomes $\chi^{-1}=\chi_{b}^{-1}-\Pi(\fvec q,\Omega_{n})$,
where $\Pi$ is the standard Lindhard function. Expanding it for small
momentum and frequency, we find:
\begin{equation}
\chi(\fvec q,\Omega_{n})=\frac{\chi_{0}}{\xi^{-2}+\fvec q^{2}+|\Omega_{n}|/\gamma}\ ,\label{eq_chi}
\end{equation}
where $\xi^{-2}=r_{0}-\chi_{0}\Pi\left(0,0\right)$ is the inverse
squared correlation length, which vanishes at the QCP, and $\gamma^{-1}=\lambda^{2}\chi_{0}/\left(2\pi v_{F}^{2}\sin\theta\right)$
is the Landau damping. Experimentally, the distance to the QCP can
be accessed by the NMR spin-lattice relaxation rate, since $T_{1}T\propto\xi^{-2}$
for a quasi-2D system. Here, $\theta$ is the angle between $\fvec v_{c}$
and $\fvec v_{d}$. To complete the model, we introduce the contributions
from the small-momentum and large-momentum impurity potentials, $u_{0}$
and $u_{Q}$ respectively:

\begin{equation}
S_{\mathrm{imp}}=\int_{kk'}u_{0}\big(c_{\fvec k\sigma}^{\dag}c_{\fvec k'\sigma}+d_{\fvec k\sigma}^{\dag}d_{\fvec k'\sigma}\big)+\int_{kk'}u_{Q}\big(c_{\fvec k\sigma}^{\dag}d_{\fvec k'\sigma}+h.c.\big)\label{S_imp}
\end{equation}

For a point-like impurity, such as considered in Ref. \cite{AVC10},
it follows that $u_{0}=u_{Q}$.

\begin{figure}[htbp]
\centering \includegraphics[width=0.9\columnwidth]{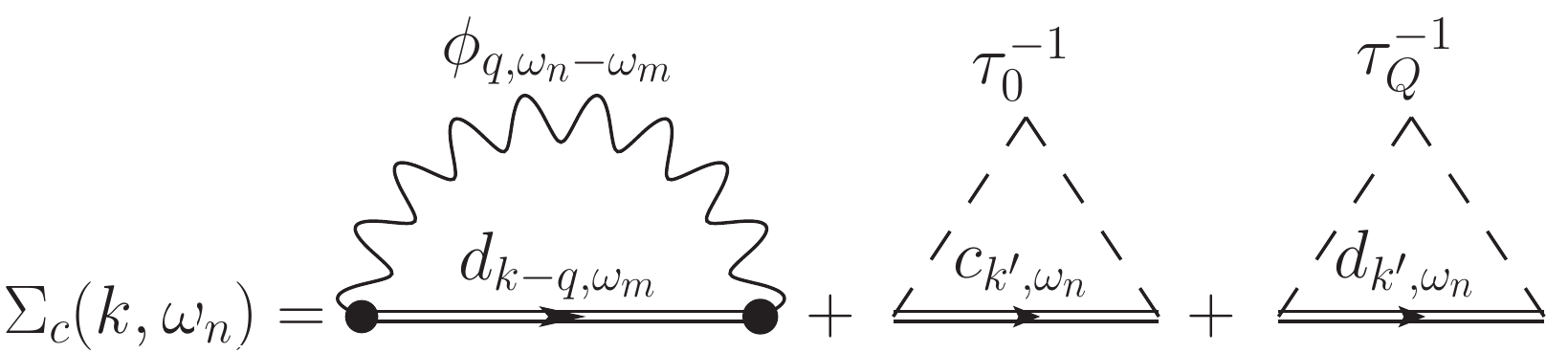} \protect\protect\caption{Feynman diagrams for the fermionic self-energy $\Sigma$, including
the fermion-boson coupling and disorder scattering.}

\label{Fig:FSE}
\end{figure}

The spin-fermion model (\ref{spin_fermion}) has been studied by a
variety of different techniques, from large-$N$\cite{AVC03,Sachdev10}
and RG \cite{SSLee15,Strack16} to Quantum Monte Carlo \cite{Berg12}.
Here, we consider the large-$N$ approach, where $N$ is the number
of hot spots. Its main advantage is that it allows one to set up an
Eliashberg-like approach to compute $T_{c}$. This is because, as
shown in Ref.~\cite{AVC03}, the vertex corrections are suppressed
by the factor $1/N$ and, thus, the SC gap equations can be obtained
by evaluating the one-loop self-energy shown in Fig.~\ref{Fig:FSE}.
The normal component of the self-energy has a real part $\Sigma'$,
which can be absorbed as a renormalization of the band dispersion,
and an imaginary part $\Sigma''$, which gives rise to a frequency-dependent
fermionic coherent spectral weight $Z_{n}^{-1}$ according to $Z_{n}=1-\Sigma''/\omega_{n}$.
The anomalous component of the self-energy, $W_{n}$, is proportional
to the frequency-dependent SC gap, $\Delta_{n}=W_{n}/Z_{n}$. Spin
fluctuations promote attraction in the SC channel in which the gap
changes sign from one hot spot to another, i.e. $W_{n}^{c}=-W_{n}^{d}\equiv W_{n}$,
corresponding to either a $d$-wave gap or an $s^{+-}$ gap, depending
on the position of the hot spots in the Brillouin zone (see Fig. \ref{Fig:FS}).
In the Nambu spinor representation, the fermionic self energy is in
the form \cite{Millis88}:
\begin{equation}
\Sigma(i\omega_{n},\fvec k)=i\omega_{n}(1-Z_{n})\sigma_{0}+\zeta\sigma_{3}+W_{n}\sigma_{1}\ ,
\end{equation}
where $\sigma_{i}$ are Pauli matrices. In principle, both $Z_{n}$
and $W_{n}$ depend on the momentum $\fvec k$. In our approach, where
only pairs of hot spots are considered, the momentum dependence is
neglected and only the frequency dependence is considered. We note
that, as discussed in Refs. \cite{Sachdev10,AVC13}, the contributions
from states beyond the hot spots can give rise to important effects.
However, in what concerns the linearized gap equations, these effects
become important when the energy scale associated with the curvature
of the Fermi surface is comparable to $T_{c}$ (see Ref. \cite{AVC13}).
Therefore, our approach is suitable for Fermi surfaces whose curvatures
are small. In this case, the fermionic Green's function is given by:
\begin{eqnarray}
 &  & G^{-1}(i\omega_{n},\fvec k)=i\omega_{n}Z_{n}-\epsilon\sigma_{3}-W_{n}\sigma_{1}\nonumber \\
 & \Longrightarrow & G(i\omega_{n},\fvec k)=-\frac{i\omega_{n}Z_{n}+\epsilon\sigma_{3}+W_{n}\sigma_{1}}{(Z_{n}\omega_{n})^{2}+\epsilon^{2}+W_{n}^{2}}\ ,
\end{eqnarray}
where we absorbed the real-part of the normal self-energy $\zeta$
in the electronic dispersion $\epsilon$.

By computing the one-loop self-energy in Fig.~\ref{Fig:FSE}, the
linearized Eliashberg equations ($T=T_{c}$) in the presence of disorder
can be written as
\begin{widetext}
\begin{align}
 & i\omega_{n}(1-Z_{n,c}(\fvec k))=3\lambda^{2}T\sum_{m}\int\frac{\rmd^{2}q}{(2\pi)^{2}}\frac{\chi_{0}}{\xi^{-2}+q^{2}+|\omega_{n}-\omega_{m}|/\gamma}\frac{-i\omega_{m}Z_{m,d}}{(\omega_{m}Z_{m,d})^{2}+\epsilon_{d}^{2}(\fvec k-\fvec q)}-i\frac{\mathrm{sgn}(\omega_{n})}{2\tau}\label{Eqn:WaveFunc}\\
 & W_{n,c}(\fvec k)=T\sum_{m}\int\frac{\rmd^{2}q}{(2\pi)^{2}}\frac{-3\lambda^{2}\chi_{0}}{\xi^{-2}+q^{2}+\frac{|\omega_{n}-\omega_{m}|}{\gamma}}\frac{W_{m,d}}{(\omega_{m}Z_{m,d})^{2}+\epsilon_{d}^{2}(\fvec k-\fvec q)}+\frac{(2\tau_{0})^{-1}W_{n,c}}{|\omega_{n}|Z_{n,c}}+\frac{(2\tau_{Q})^{-1}W_{n,d}}{|\omega_{n}|Z_{n,d}}\label{Eqn:Pairing}
\end{align}

\end{widetext}

The subscripts $c$, $d$ refer to the fermionic states around the
two hot spots. The equations for $Z_{n,d}$ and $W_{n,d}$ assume
similar forms. The total impurity scattering rate is given by $\tau^{-1}=\tau_{0}^{-1}+\tau_{Q}^{-1}$,
where $\tau_{0}^{-1}=2\pi n_{\mathrm{imp}}u_{0}^{2}N_{f}$ is the
small-momentum scattering rate and $\tau_{Q}^{-1}=2\pi n_{\mathrm{imp}}u_{Q}^{2}N_{f}$
is the large-momentum scattering rate, with $n_{\mathrm{imp}}$ denoting
the concentration of impurities and $N_{f}$, the density of states
at the Fermi level. Both Eqs. (\ref{Eqn:WaveFunc}) and (\ref{Eqn:Pairing})
contain the two-dimensional integral over momenta $q_{\parallel}$
and $q_{\perp}$, i.e. the components of $\fvec q$ parallel and perpendicular
to the Fermi surface. Focusing at the hot spot ($\fvec k=0$), the
fermionic self-energy is
\begin{align}
 & \int\frac{\rmd^{2}q}{(2\pi)^{2}}\frac{\chi_{0}}{\xi^{-2}+q^{2}+|\omega_{n}-\omega_{m}|/\gamma}\frac{-i\omega_{m}Z_{m,d}}{(\omega_{m}Z_{m,d})^{2}+\epsilon_{d}^{2}(\fvec q)}\nonumber \\
= & \int\frac{\rmd q_{\perp}}{(2\pi)}\frac{\chi_{0}/2}{\sqrt{\xi^{-2}+q_{\perp}^{2}+|\omega_{n}-\omega_{m}|/\gamma}}\frac{-i\omega_{m}Z_{m,d}}{(\omega_{m}Z_{m,d})^{2}+(v_{f}q_{\perp})^{2}}\nonumber \\
\approx & \frac{\chi_{0}}{4v_{f}}\frac{-i\mathrm{sgn}(\omega_{m})}{\sqrt{\xi^{-2}+|\omega_{n}-\omega_{m}|/\gamma}}\ .\label{Eqn:A}
\end{align}
where, in the last step, we considered that:
\begin{equation}
\xi^{-2}+|\omega_{n}-\omega_{m}|/\gamma\gg\left(\frac{\omega_{m}Z_{m}}{v_{f}}\right)^{2}\label{Eqn:Cutoff}
\end{equation}
which naturally establishes a cutoff:
\begin{equation}
\Omega_{c}=\max\left(\frac{v_{f}^{2}}{\gamma},\frac{v_{f}}{\xi}\right)=\max\left(\frac{8\Lambda}{9\pi\sin^{2}\theta},\frac{4\sqrt{r\Lambda}}{3\sin\theta}\right)\label{Eqn:CutoffAux}
\end{equation}
where $r=\frac{\xi^{-2}\gamma}{2\pi}$ is the energy scale of the
AFM fluctuations, and $\Lambda=\frac{9}{16}\lambda^{2}\chi_{0}\sin\theta$
is an effective coupling constant. Note that this cutoff arises not
from the bandwidth, but from the restriction in the momentum integration.
For notation convenience, we define
\begin{equation}
A(\omega_{n}-\omega_{m})=\frac{\chi_{0}}{2\sqrt{\xi^{-2}+|\omega_{n}-\omega_{m}|/\gamma}}
\end{equation}
Therefore, the Eliashberg equations are given by:
\begin{eqnarray}
Z_{n} & = & 1+\frac{3\lambda^{2}T}{2v_{f}\omega_{n}}\sum_{m}\mathrm{sgn}(\omega_{m})A(\omega_{n}-\omega_{m})+\frac{\tau_{0}^{-1}+\tau_{Q}^{-1}}{2|\omega_{n}|}\label{Eqn:EliashbergZ}\\
W_{n} & = & \frac{3\lambda^{2}T}{2v_{f}}\sum_{m}\frac{W_{m}}{Z_{m}}A(\omega_{n}-\omega_{m})+\frac{W_{n}\left(\tau_{0}^{-1}-\tau_{Q}^{-1}\right)}{2|\omega_{n}|Z_{n}}\label{Eqn:EliashbergGap}
\end{eqnarray}

Our goal is to investigate how $\rmd T_{c}/\rmd\tau^{-1}$ deviates
from the universal AG result, $\left(\rmd T_{c}/\rmd\tau^{-1}\right)_{\mathrm{AG}}=-\pi/4$.
To gain insight into this problem, we reexpress the Eliashberg equations
as a functional form \cite{BRMethod,Valls78,Millis88,RMF13}. In particular,
after defining $\bar{\Delta}_{n}=TW_{n}/(Z_{n}|\omega_{n}|)$ and
restricting the solution to even-frequency pairing, $W\left(-\omega_{n}\right)=W\left(\omega_{n}\right)$,
solving the Eliashberg equations becomes equivalent to finding the
zero eigenvalue $\eta$ of $\hat{K}_{mn}\bar{\Delta}_{n}=\eta\bar{\Delta}_{m}$.
Here, the matrix $\hat K$ is given by:
\begin{eqnarray}
\hat{K}_{m\neq n} & = & \sqrt{\frac{\Lambda}T} \left( \frac1{\sqrt{|m - n| + r/T}} + \frac1{\sqrt{m + n + 1 + r/T}} \right) \nonumber \\
\hat{K}_{nn} & = & \sqrt{\frac{\Lambda}T} \frac1{\sqrt{2n+1 + r/T}} - \pi(2n + 1) \nonumber \\
& & - \sqrt{\frac{\Lambda}T} \sum_{n' \neq n} \frac{\mathrm{sgn}(\omega_{n'})}{\sqrt{|n - n'| + r/T}} - \frac{\tau_Q^{-1}}T
\label{Eq:TcMatrix}
\end{eqnarray}
where $m,n$ are non-negative integers. $T_{c}$ is obtained when
the largest eigenvalue $\eta$ vanishes. These equations reduce to
those studied in Ref.~\cite{AVC10} when $\tau_{0}^{-1}=\tau_{Q}^{-1}$.
The main advantage of this functional approach is that it allows us
to study the impact of weak disorder on $T_{c}$ without having to
solve explicitly the disordered equations. This is accomplished by
employing the Hellmann-Feynman theorem:

\begin{equation}
\frac{\rmd T_{c}}{\rmd\tau^{-1}}=-\left\langle \frac{\rmd\hat{K}}{\rmd\tau^{-1}}\right\rangle /\left\langle \frac{\rmd\hat{K}}{\rmd T_{c}}\right\rangle \equiv-\left(\frac{\rmd\eta}{\rmd\tau^{-1}}\right)/\left(\frac{\rmd\eta}{\rmd T_{c}}\right)\label{Hellmann_Feynman}
\end{equation}
where $\left\langle \cdots\right\rangle $ refers to an average with
respect to the normalized eigenvector $\bar{\Delta}_{n}$ of the system
without disorder and $\eta=\sum_{m,n}\hat{K}_{mn}\bar{\Delta}_{m}\bar{\Delta}_{n}$.
Next, we divide the contributions to $\left(\frac{\rmd T_{c}}{\rmd\tau^{-1}}\right)$
in two classes: those arising from the coupling between disorder and
the fermionic degrees of freedom, $\left(\frac{\rmd T_{c}}{\rmd\tau^{-1}}\right)_{f,i}$,
and those arising from the coupling between disorder and the bosonic
degrees of freedom (i.e. the pairing interaction), $\left(\frac{\rmd T_{c}}{\rmd\tau^{-1}}\right)_{b,j}$.
While the former corresponds simply to the $\tau_{Q}^{-1}$ term in
Eq. (\ref{Eq:TcMatrix}), the latter is implicit in the kernel (\ref{Eq:TcMatrix})
via the dependence of the pairing interaction $A(\Omega_{n})$ on
disorder. Because of Hellmann-Feynman theorem, these contributions
can be treated independently and just added up in the end:

\begin{equation}
\frac{\rmd T_{c}}{\rmd\tau^{-1}}=\sum_{i}\left(\frac{\rmd T_{c}}{\rmd\tau^{-1}}\right)_{f,i}+\sum_{j}\left(\frac{\rmd T_{c}}{\rmd\tau^{-1}}\right)_{b,j}\label{main-1}
\end{equation}

\section{Suppression of $T_{c}$ due to the coupling of disorder and the electronic
degrees of freedom}

We first investigate how the coupling between disorder and the fermionic
states affects the suppression rate $\left(\frac{\rmd T_{c}}{\rmd\tau^{-1}}\right)$.
As it is immediate clear from Eq. (\ref{Eq:TcMatrix}), there is only
one term in the kernel that depends on the impurity scattering explicitly,
giving rise to the contribution $\left(\frac{\rmd T_{c}}{\rmd\tau^{-1}}\right)_{f}$.
In particular, because only the large-momentum scattering rate $\tau_{Q}^{-1}$
appears in the functional $\hat{K}$, $T_{c}$ is insensitive to small-momentum
scattering $\tau_{0}^{-1}$ -- an extension of the Anderson theorem
to sign-changing SC gaps. Before we numetically evaluate (\ref{Hellmann_Feynman}),
it is instructive to consider two limiting cases: the BCS limit and
quantum critical pairing.

\subsection{BCS Limit}

The BCS limit corresponds to the case in which the system is far away
from the QCP and the coupling constant is small, $r\gg\Omega_{c}\gg\Lambda$.
The pairing interaction then becomes frequency-independent and small,
$A(\Omega_{n})\propto r^{-1/2}$, and the fermionic coherence factor
can be approximated by $Z_{n}\approx1$. In this limit, Eq.~(\ref{Eq:TcMatrix})
becomes
\begin{align}
K_{mn}\approx & 2\sqrt{\frac{\Lambda}{r}}-\pi(2n+1)\delta_{mn}\qquad\Longrightarrow\qquad\nonumber \\
\bar{\Delta}_{n\geq0}= & \frac{2}{\pi(2n+1)}\sqrt{\frac{\Lambda}{r}}\sum_{m\leq\frac{\Omega_{c}}{2\pi T}}\bar{\Delta}_{m}\label{Eq:BCS}
\end{align}
Defining the quantity $c=\sum_{m}\bar{\Delta}_{m}$, we obtain the
self-consistent equation:
\begin{align}
 & c=\frac{c}{\pi}\sqrt{\frac{\Lambda}{r}}\sum_{n\leq\Omega_{c}/(2\pi T)}\frac{1}{n+1/2}\nonumber \\
 & \quad=\frac{c}{\pi}\sqrt{\frac{\Lambda}{r}}\left[\psi\left(\frac{\Omega_{c}}{2\pi T}-\frac{1}{2}\right)-\psi\left(\half\right)\right]\nonumber \\
\Longrightarrow\quad & T_{c}\approx\frac{\Omega_{c}}{2\pi}\exp\left[-\pi\sqrt{\frac{r}{\Lambda}}-\psi\left(\half\right)\right]
\end{align}
which agrees with the standard BCS expression. Here, $\psi\left(x\right)$
is the digamma function.

As shown in Eq.~(\ref{Eq:BCS}), the matrix elements of $\hat{K}$
are independent of $T$, but the eigenvalue $\eta$ still depends
on $T$ via the changes in the matrix size $N_{c}$, which is set
by the hard cutoff $\Omega_{c}$ via $N_{c}=\Omega_{c}/(2\pi T)$.
To take this effect into account, consider a reduction in the matrix
size by $1$, $N_{c}\rightarrow N_{c}-1$, which means that the last
row and the last column no longer take part in the determination of
the eigenvalue. Then, the change in $\eta=\sum_{m,n}\hat{K}_{mn}\bar{\Delta}_{m}\bar{\Delta}_{n}$
is given by:
\begin{align}
\delta\eta & =-\sum_{m}\left(\hat{K}_{mN_{c}}+\hat{K}_{N_{c}m}\right)\bar{\Delta}_{m}\bar{\Delta}_{N_{c}}+\hat{K}_{N_{c}N_{c}}\left(\bar{\Delta}_{N_{c}}\right)^{2}\nonumber \\
 & =\hat{K}_{N_{c}N_{c}}\left(\bar{\Delta}_{N_{c}}\right)^{2}
\end{align}
Therefore, we find:
\begin{align}
\frac{\delta\eta}{\delta T_{c}}= & -\frac{\Omega_{c}}{2\pi T^{2}}\frac{\delta\eta}{\delta N_{c}}=\frac{\Omega_{c}}{2\pi T^{2}}\hat{K}_{N_{c}N_{c}}\left(\bar{\Delta}_{N_{c}}\right)^{2}
\end{align}
yielding:

\begin{align}
\frac{\rmd T_{c}}{\rmd\tau_{Q}^{-1}}= & -\left(\frac{\rmd\eta}{\rmd T_{c}}\right)^{-1}\frac{\rmd\eta}{\rmd\tau_{Q}^{-1}}\nonumber \\
= & \frac{1}{N_{c}}\frac{1}{\hat{K}_{N_{c}N_{c}}\left(\bar{\Delta}_{N_{c}}\right)^{2}}
\end{align}

Using Eq. (\ref{Eq:BCS}), we have $\hat{K}_{N_{c}N_{c}}=-\pi(2N_{c}+1)\approx-2\pi N_{c}$.
Furthermore, from the same equation, we have:
\begin{equation}
\bar{\Delta}_{N_{c}}=\frac{c}{\pi(N_{c}+\frac{1}{2})}\sqrt{\frac{\Lambda}{r}}
\end{equation}
where $c=\sum_{m\leq\frac{\Omega_{c}}{2\pi T}}\bar{\Delta}_{m}$.
The value of $c$ can be obtained by normalizing the eigenvector:
\begin{equation}
\sum_{n=0}^{\frac{\Omega_{c}}{2\pi T}}\bar{\Delta}_{n}^{2}=1\quad\Longrightarrow\quad\frac{c}{\pi}\sqrt{\frac{\Lambda}{r}}=\left(\sum_{n}\frac{1}{\left(n+1/2\right)^{2}}\right)^{-1/2}
\end{equation}

Therefore:
\begin{align}
 & \bar{\Delta}_{N_{c}}=\frac{1}{N_{c}+1/2}\left(\sum_{n}\frac{1}{(n+1/2)^{2}}\right)^{-1/2}\approx\frac{\sqrt{2}}{\pi N_{c}}\nonumber \\
\quad\Longrightarrow\quad & \frac{\rmd T_{c}}{\rmd\tau_{Q}^{-1}}\approx-\frac{1}{N_{c}}\frac{1}{2\pi N_{c}}\left(\frac{\pi N_{c}}{\sqrt{2}}\right)^{2}=-\frac{\pi}{4}\ .
\end{align}
recovering the Abrikosov-Gor'kov universal value for dirty superconductors.

\subsection{Quantum Critical Pairing Limit}

The second limiting case corresponds to the system at the QCP, for
which $r\propto\xi^{-2}=0$. In this case, the pairing interaction
is strongly frequency-dependent, $A\left(\Omega_{n}\right)\propto\Omega_{n}^{-1/2}$.
From Eq.~(\ref{Eqn:EliashbergZ}), we find that as $T\rightarrow0$
the low-frequency coherent factor vanishes as $Z^{-1}(\omega\ll\Omega_{c})\propto\omega^{1/2}$,
a hallmark of non-Fermi liquid behavior. An interesting property of
the system of equations at the QCP is that they converge in the limit
$\Omega_{c}\rightarrow\infty$, i.e. $T_{c}$ and $\left(\frac{\rmd T_{c}}{\rmd\tau^{-1}}\right)$
do not depend on the cutoff. In this limit, the $\hat{K}$ matrix
becomes:
\begin{eqnarray}
\hat{K}_{nn} & = & \frac{\sqrt{\Lambda/T}}{\sqrt{2n+1}}-\pi(2n+1)-\frac{1}{\tau_{Q}T}\nonumber \\
 &  & -2\sqrt{\frac{\Lambda}{T}}\left[\zeta\left(\half,1\right)-\zeta\left(\half,n+1\right)\right]\\
\hat{K}_{m\neq n} & = & \sqrt{\frac{\Lambda}{T}}\left(\frac{1}{\sqrt{|n-m|}}+\frac{1}{\sqrt{n+m+1}}\right)\ .
\end{eqnarray}
where $\zeta\left(\half,x\right)$ is the Hurwitz zeta function. Clearly,
the only free parameter is the combination $\frac{\Lambda}{T_{c}}$.
By numerically diagonalizing the matrix, we find $T_{c}\approx0.5\Lambda$.
Analysis of the eigenvalue problem for large frequencies reveal that
$\bar{\Delta}_{n}\propto n^{-3/2}$, which explains why the problem
converges for $\Omega_{c}\rightarrow\infty$. This is to be contrasted
with the BCS case, in which $\bar{\Delta}_{n}\propto n^{-1}$, implying
that the sum does not converge and a cutoff is needed.

To compute $\left(\frac{\rmd T_{c}}{\rmd\tau^{-1}}\right)$, we use
the Hellmann-Feynman theorem, Eq. (\ref{Hellmann_Feynman}). Using
the equations above, we find:

\begin{equation}
\frac{\partial\hat{K}_{mn}}{\partial\tau_{0}^{-1}}=0\qquad\frac{\partial\hat{K}_{mn}}{\partial\tau_{Q}^{-1}}=-\delta_{mn}/T
\end{equation}
and:

\begin{equation}
\frac{\pd\hat{K}_{mn}}{\pd T}=-\frac{\hat{K}_{mn}}{2T}-\frac{\pi(2n+1)}{2T}\,\delta_{mn}
\end{equation}

Hence, we obtain:
\begin{eqnarray}
 &  & \left(\frac{\dif\eta}{\rmd T}\right)=-\frac{\pi}{2T}\sum_{n=0}^{\infty}\bar{\Delta}_{n}^{2}(2n+1)\nonumber \\
 & \Longrightarrow & \frac{\rmd T_{c}}{\rmd\tau_{Q}^{-1}}=-\frac{2}{\pi}\left(\sum_{n=0}^{\infty}\bar{\Delta}_{n}^{2}(2n+1)\right)^{-1}\label{Eqn:Delta_dT}
\end{eqnarray}

The term inside the brackets does not depend on any free parameters,
and therefore can be evaluated numerically. Numerical evaluation gives
$\rmd T_{c}/\rmd\tau_{Q}^{-1}\approx-0.45$, which is smaller than
the AG universal value obtained away from the QCP.

\subsection{General case}

\begin{figure}[htbp]
\begin{centering}
\subfigure[\label{Fig:TcInf:Tc}]{\includegraphics[width=0.8\columnwidth]{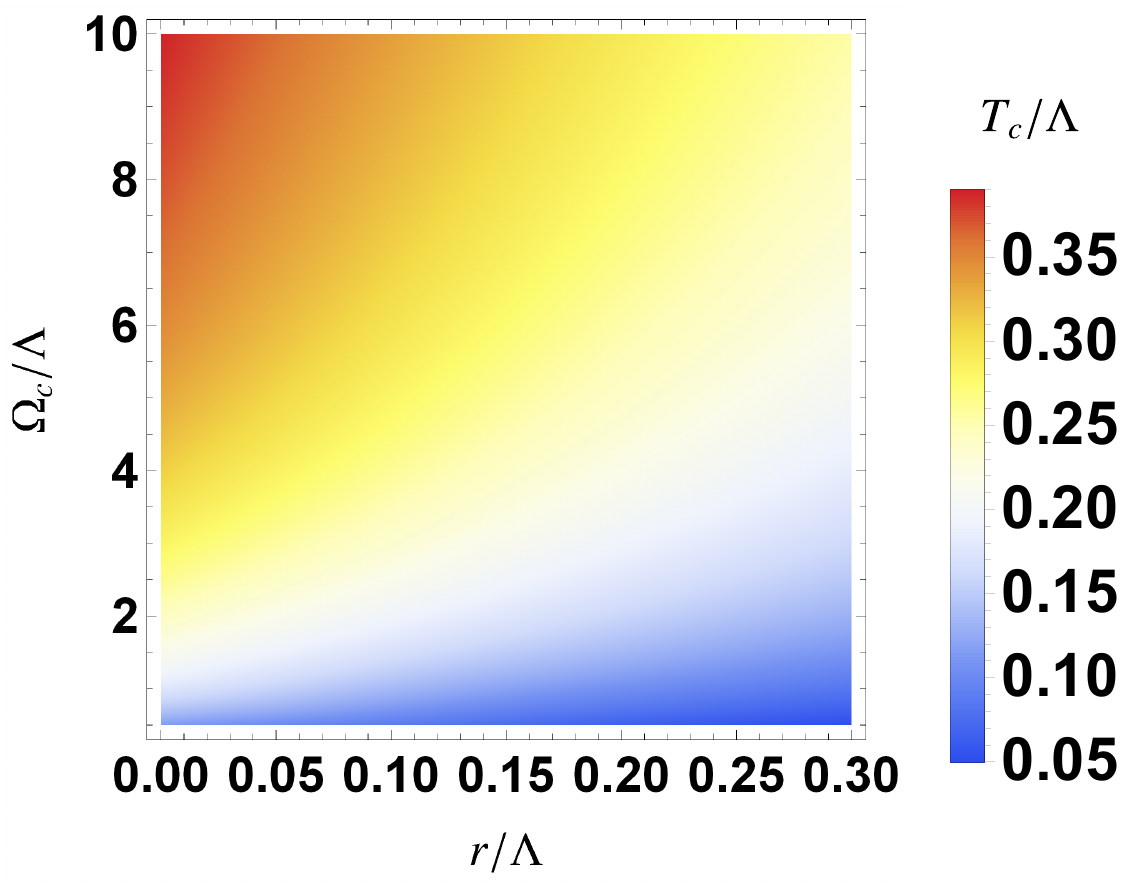}}
\subfigure[\label{Fig:TcInf:TcInter}]{\includegraphics[width=0.8\columnwidth]{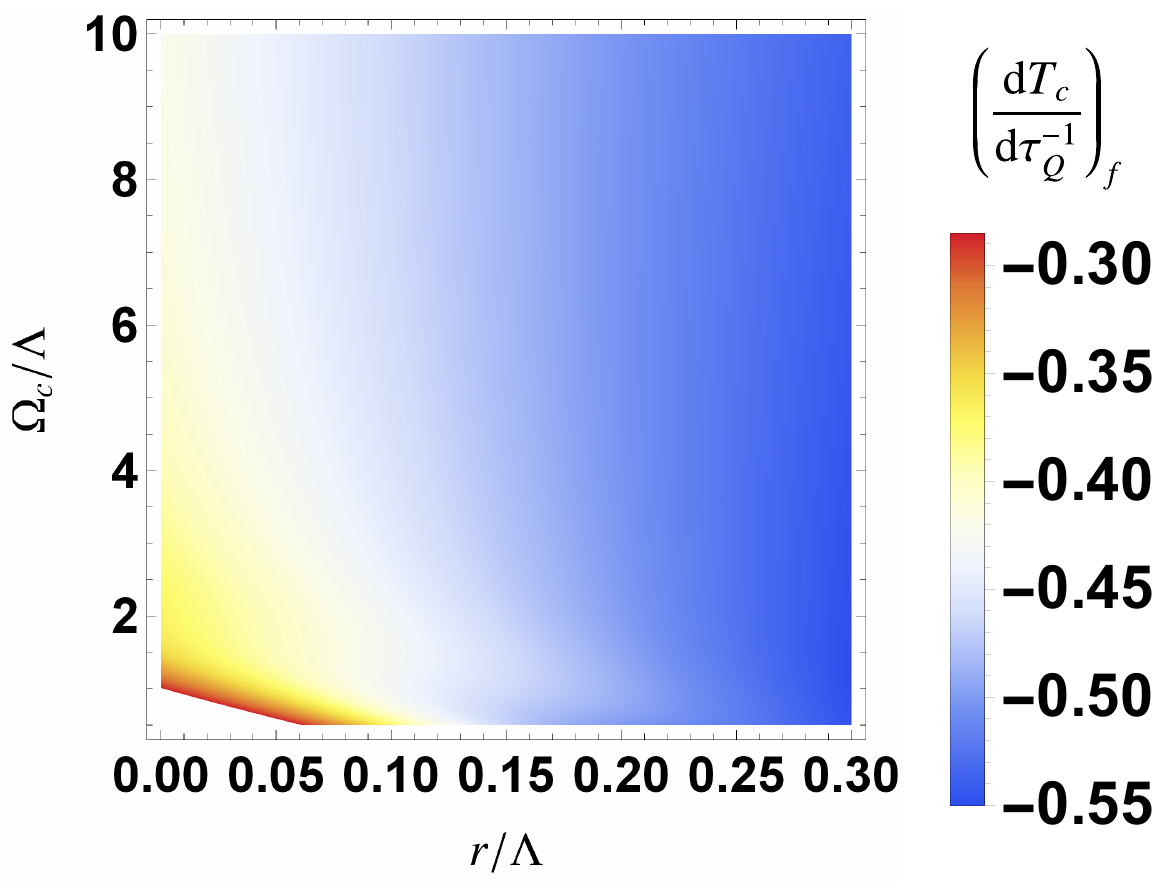}}
\protect\protect\caption{ (a) Transition temperature of the clean system ($T_{c}$) as a function
of the distance to the QCP $r$ and the cutoff $\Omega_{c}$ (in units
of the effective coupling $\Lambda$). (b) The suppression of $T_{c}$
by disorder, $dT_{c}/d\tau^{-1}$, when the system is near the quantum
critical point $r=0$. Only the effects of the coupling of disorder
to the fermionic degrees of freedom are included. In the plot, we
used the soft cutoff procedure. }

\par\end{centering}

\label{Fig:TcInf}
\end{figure}

Therefore, the two limiting cases ($r=0$ and $r\gg\Omega_{c}\gg\Lambda$)
suggest that proximity to the QCP promotes the robustness of $T_{c}$
against pair-breaking disorder. Before presenting the results for
a general distance $r$ from the QCP, we first explain how the high-energy
cutoff $\Omega_{c}$ is set in our calculation. In the subsection
discussing the BCS limit, we set a hard cutoff $N_{c}=\Omega_{c}/\left(2\pi T\right)$
in the Matsubara sum. Although this procedure does not affect the
behavior of $T_{c}$ in the limit of $\Omega_{c}\gg T_{c}$, it will
make $T_{c}$ behave discontinuously as $\Omega_{c}$ and $r$ decrease.
To avoid such a discontinuity, we set instead a soft cutoff in the
Matsubara sum by including an appropriate continuous function $f\left(\omega_{n}\right)$
that is strongly suppressed above $\Omega_{c}$ and nearly $1$ below
$\Omega_{c}$. Specifically, we change the bosonic propagator to:
\begin{eqnarray}
 &  & \frac{1}{|\omega_{n}-\omega_{m}|/\gamma+\fvec q^{2}+\xi^{-2}}\rightarrow\frac{f(\omega_{n})f(\omega_{m})}{|\omega_{n}-\omega_{m}|/\gamma+\fvec q^{2}+\xi^{-2}}
\end{eqnarray}
with $f(\omega)=\left[\exp\left(\frac{|\omega|-\Omega_{c}}{\Lambda_{d}}\right)+1\right]^{-1}$,
where $\Lambda_{d}=0$ gives the hard energy cutoff. This function
has the property that when $|\omega|\ll\Omega_{c}$, $f(\omega)\approx1$,
and when $|\omega|\gg\Omega_{c}$, $f(\omega)\approx0$. In our calculation,
$\Lambda_{d}$ is set to be $\Lambda_{d}=\max\left(0.1\Lambda,0.3\Omega_{c}\right)$.
We emphasize that none of our results qualitatively change for $\Lambda_{d}=0$
or $\Lambda_{d}$ small. Thus, at the QCP, the matrix $\hat{K}$ with
such soft high-energy cutoff becomes:

\begin{align}
\hat{K}_{m\neq n}= & \sqrt{\frac{\Lambda}{T}}\left(\frac{f_{m}f_{n}}{\sqrt{|n-m|+r/T}}+\frac{f_{m}f_{n}}{\sqrt{n+m+1+r/T}}\right)\label{Eqn:KOffDia}\\
\hat{K}_{nn}= & f_{n}^{2}\sqrt{\frac{\Lambda}{(2n+1)T+r}}-\pi(2n+1)\nonumber \\
 & -f_{n}\sqrt{\frac{\Lambda}{T}}\sum_{m\neq n}\frac{\mathrm{sgn}(\omega_{m})f_{m}}{\sqrt{|n-m|+r/T}}-\frac{1}{\tau_{Q}T}
\label{Eqn:KDia}
\end{align}
where

\begin{equation}
f_{n\geq0}=f\left(\pi(2n+1)T\right)\label{Eqn:WeightFunc}
\end{equation}
is the weight function. In the clean limit ($\tau_{Q}^{-1}=0$), it
is straightforward to compute $T_{c}$ by diagonalizing the matrix
$\hat{K}$. As show in Fig.~\ref{Fig:TcInf}, we find that $T_{c}$
is generally suppressed away from the QCP for a fixed value of the
cutoff $\Omega_{c}$.

We are now in position to compute the suppression rate of $T_{c}$
by disorder using the Hellmann-Feynman theorem, Eq. (\ref{Hellmann_Feynman}),
combined with the solution of the clean system discussed above. From
Eqs.~(\ref{Eqn:KOffDia}) and (\ref{Eqn:KDia}), we have $\partial\hat{K}/\partial\tau_{0}^{-1}=0$,
and $\partial\hat{K}/\partial\tau_{Q}^{-1}=-1/T$, i.e. only large-momentum
scattering is pair-breaking. Since we apply the soft cutoff here,
the weight function $f$ also depends on temperature, as shown by
Eq.~(\ref{Eqn:WeightFunc}). We have:
\begin{widetext}
\begin{align}
\frac{\pd\hat{K}_{m\neq n}}{\pd T}= & -\frac{\pi}{\Lambda_{d}}\sqrt{\frac{\Lambda}{T}}f_{m}f_{n}\Big((2n+1)(1-f_{n})+(2m+1)(1-f_{m})\Big)\left(\frac{1}{\sqrt{|n-m|+r/T}}+\frac{1}{\sqrt{n+m+1+r/T}}\right)\nonumber \\
 & +\frac{r}{2T^{2}}\sqrt{\frac{\Lambda}{T}}f_{m}f_{n}\left[\left(|n-m|+r/T\right)^{-\frac{3}{2}}+\left(n+m+1+r/T\right)^{-\frac{3}{2}}\right]-\frac{\hat{K}_{mn}}{2T}\nonumber \\
\frac{\pd\hat{K}_{nn}}{\pd T}= & \frac{\pi}{\Lambda_{d}}\sqrt{\frac{\Lambda}{T}}f_{n}\sum_{m\neq n}\frac{\mathrm{sgn}(\omega_{m})f_{m}\Big(|2m+1|(1-f_{m})+(2n+1)(1-f_{n})\Big)}{\sqrt{|n-m|+r/T}}-\frac{r}{2T^{2}}\sqrt{\frac{\Lambda}{T}}f_{n}\sum_{m\neq n}\frac{\mathrm{sgn}(\omega_{m})f_{m}}{\left(|n-m|+r/T\right)^{3/2}}\nonumber \\
 & -\frac{\hat{K}_{nn}}{2T}-\frac{\pi(2n+1)}{2T}-\frac{2\pi(2n+1)}{\Lambda_{d}}\sqrt{\frac{\Lambda}{T}}\frac{f_{n}^{2}(1-f_{n})}{\sqrt{2n+1+r/T}}+\frac{r}{2T^{2}}\sqrt{\frac{\Lambda}{T}}\frac{f_{n}^{2}}{(2n+1+r/T)^{3/2}}\label{Eqn:TDerivative}
\end{align}

\end{widetext}

Calculating these expressions, we present in Fig.~\ref{Fig:TcInf}
$\rmd T_{c}/\rmd\tau_{Q}^{-1}$ in the proximity of a QCP. The results
agree with our expectations and reveal that $T_{c}$ is indeed in
general more robust against disorder at the QCP ($r=0$), specially
when compared to the AG universal value $-\pi/4\approx-0.785$. Although
the precise values for $T_{c}$ and $-\rmd T_{c}/\rmd\tau_{Q}^{-1}$
depend on the ratio $\Omega_{c}/\Lambda$, the general trend is robust,
and $-\rmd T_{c}/\rmd\tau_{Q}^{-1}$ remains well below the AG universal
value $\pi/4$ as shown in Fig~\ref{Fig:QCPSC}. To understand this
behavior, we note that the last term of the Eliashberg equation (\ref{Eqn:EliashbergGap}),
proportional to $\tau_{Q}^{-1}$, effectively reduces the pairing
vertex to $W_{n}\rightarrow W_{n}/\left(1+\frac{\tau_{Q}^{-1}}{2Z_{n}|\omega_{n}|}\right)$.
Therefore, because at the QCP the fermionic coherent weight $Z^{-1}\propto\omega^{1/2}$
vanishes at the Fermi surface, the effect of disorder on the pairing
vertex becomes less relevant at low frequencies, where the gap function
is the largest. As the system moves away from the QCP, $Z^{-1}$ enhances
at the Fermi level, and disorder becomes more relevant.

\section{Suppression of $T_{c}$ due to the coupling of disorder and bosonic
degrees of freedom}

Our analysis so far agrees with the general results from Ref.~\cite{AVC10}
and mirrors the standard AG approach for conventional dirty superconductors,
with disorder impacting the electronic degrees of freedom. In this
regard, one of the main differences between the conventional and unconventional
SC cases stems from the reduced coherent electronic spectral weight
near the QCP. There is however another important difference between
the two cases: while in the former the pairing interaction arises
from an independent degree of freedom (phonons), in the latter it
arises from the same electronic degrees of freedom (AFM fluctuations).
Since disorder affects the fermionic states, it must then change also
the AFM fluctuation spectrum.

\begin{figure}[htbp]
\begin{centering}
\subfigure{\label{Fig:Boson:DerBoson}}{\includegraphics[width=0.95\columnwidth]{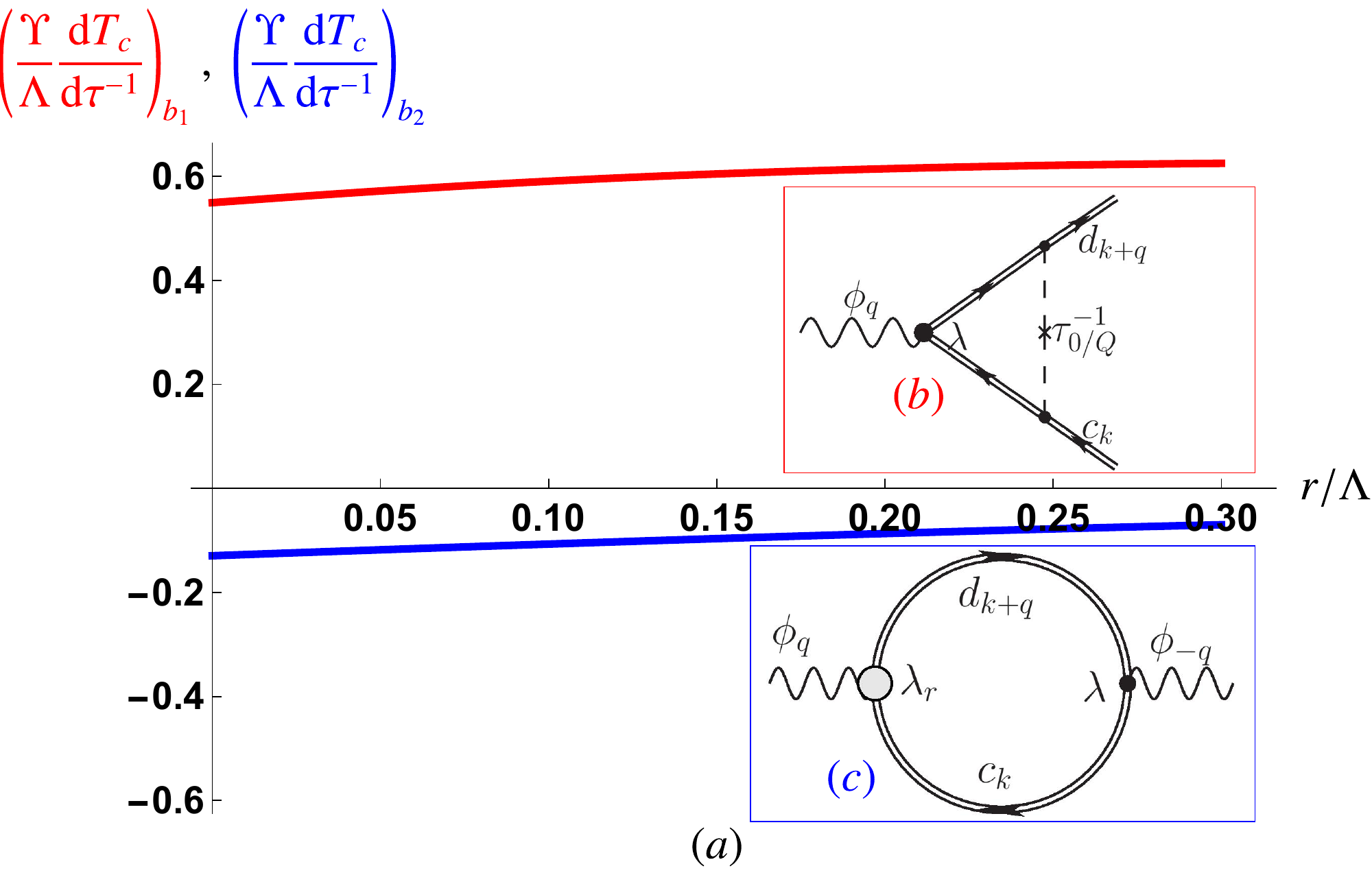}}
\addtocounter{subfigure}{2} \subfigure[\label{Fig:Boson:TcIntra}]{\includegraphics[width=0.8\columnwidth]{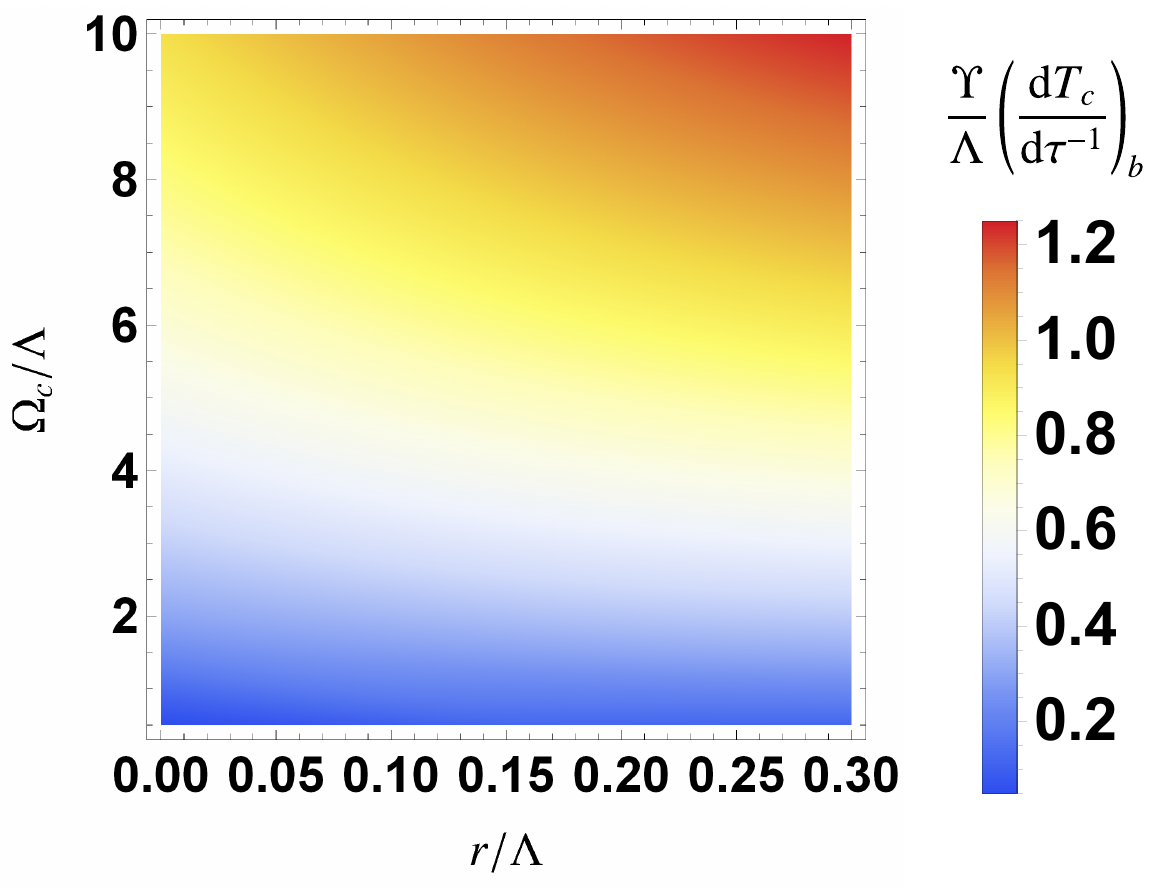}}
\protect\protect\caption{(a) Two contributions to the derivative of $T_{c}$ with respect to
the total scattering rate $\tau^{-1}=\tau_{0}^{-1}+\tau_{Q}^{-1}$
that arise from the coupling between disorder and bosonic degrees
of freedom. The positive contribution $b_{1}$ corresponds to the
dressing of the fermion-boson vertex by disorder (inset b), whereas
the negative contribution $b_{2}$ corresponds to the dressing of
the bosonic self-energy by disorder (inset c). (d) The change of $T_{c}$
due to the coupling of disorder to the bosons, $\left(dT_{c}/d\tau^{-1}\right)_{b}=\left(dT_{c}/d\tau^{-1}\right)_{b,1}+\left(dT_{c}/d\tau^{-1}\right)_{b,2}$,
when the system is around the QCP $r=0$. }

\par\end{centering}

\label{Fig:Boson}
\end{figure}

Within our functional approach to the Eliashberg equations, including
this effect is straightforward within linear order in $\tau^{-1}$.
Specifically, we need to compute how the pairing interaction, as defined
in (\ref{Eqn:A}), changes in the presence of disorder. We identify
two processes through which the bosonic degrees of freedom are affected
by disorder scattering, as shown in Fig.~\ref{Fig:Boson} (b) and
(c) (see also Ref. \cite{Strinati13}). The first process, Fig.~\ref{Fig:Boson}
(b), corresponds to the renormalization of the electron-boson vertex
by disorder, and gives rise to the contribution $\left(\frac{\rmd T_{c}}{\rmd\tau^{-1}}\right)_{b,1}$
for the suppression of $T_{c}$. The second process, Fig.~\ref{Fig:Boson}
(c), corresponds to the renormalization of the bosonic self-energy,
and gives rise to the contribution $\left(\frac{\rmd T_{c}}{\rmd\tau^{-1}}\right)_{b,2}$.
In the spirit of the large-$N$ expansion, the renormalization of
the disorder vertex by the bosonic fluctuations is small by a $1/N$
factor, and therefore will not be considered hereafter.

\subsection{Electron-boson vertex renormalization}

We first calculate how the electron-boson vertex is renormalized by
disorder. As shown in Fig.~\ref{Fig:Boson}(b), the vertex correction
$\delta\lambda$ is given by:
\begin{align}
\delta\lambda= & n_{\mathrm{imp}}u^{2}\int\frac{\rmd^{2}k}{(2\pi)^{2}}G_{c}\left(i\omega_{n},\fvec k-\frac{\fvec q}{2}\right)G_{d}\left(i\omega_{m},\fvec k+\frac{\fvec q}{2}+\fvec Q\right)\nonumber \\
= & \frac{n_{\mathrm{imp}}u^{2}}{|\fvec v_{c}\times\fvec v_{d}|}\int\frac{\rmd\epsilon_{1}\rmd\epsilon_{2}}{(2\pi)^{2}}\frac{1}{i\omega_{n}Z_{n}+\fvec v_{c}\cdot\fvec q/2-\epsilon_{1}}\times\nonumber \\
 & \qquad\qquad\quad\frac{1}{i\omega_{m}Z_{m}-\fvec v_{c}\cdot\fvec q/2-\epsilon_{2}}
\end{align}
where $n_{\mathrm{imp}}$ is the impurity concentration and $u$ is
the impurity potential. Since the effect is the same for small and
large momentum scattering, we do not distinguish them here. Using
the result:
\begin{eqnarray}
 &  & \int_{-\infty}^{\infty}\frac{\rmd p}{ia-c-p}=-i\pi\mathrm{sgn}(a)\ .
\end{eqnarray}
we obtain:
\begin{align}
\delta\lambda=-\lambda\frac{\mathrm{sgn}(\omega_{n}\omega_{m})}{8\pi N_{f}\tau|\fvec v_{c}\times\fvec v_{d}|}\ .\label{delta_lambda}
\end{align}

Note that the vertex correction depends only on the external frequencies
of the two fermion legs. In the ladder approximation, we find the
renormalized vertex $\lambda_{r}$
\begin{align}
 & \lambda_{r}=\lambda\left(1+\frac{\mathrm{sgn}(\omega_{n}\omega_{m})}{8\pi N_{f}\tau|\fvec v_{c}\times\fvec v_{d}|}\right)^{-1}\nonumber \\
 & \Longrightarrow\left.\frac{\rmd\lambda_{r}}{\rmd\tau^{-1}}\right|_{\tau^{-1}=0}=-\lambda\frac{\mathrm{sgn}(\omega_{n}\omega_{m})}{2\Upsilon}\ .\label{Eqn:lambdaSign}
\end{align}
where we defined the energy scale $\Upsilon=4\pi N_{f}v_{f}^{2}\sin\theta$.
Thus, we have two different behaviors depending of whether the external
frequencies have the same sign ($\lambda_{+}$) or different signs
($\lambda_{-}$)~\cite{Jorg12}:
\begin{equation}
\lambda_{\pm}=\lambda\left(1\pm\frac{1}{2\Upsilon\tau}\right)^{-1}\ ,\quad\left.\frac{\rmd\lambda_{\pm}}{\rmd\tau^{-1}}\right|_{\tau^{-1}=0}=\mp\frac{\lambda}{2\Upsilon}\label{Eqn:lambda}
\end{equation}

\subsection{Bosonic self-energy renormalization}

We now calculate the renormalization of the particle-hole bubble by
disorder. As shown in Fig.~\ref{Fig:Boson}(c), we have:
\begin{align}
\Pi(i\Omega_{m},\fvec q)= & -2T\sum_{n}\int\frac{\rmd^{2}q}{(2\pi)^{2}}\lambda_{r}\lambda G_{c}\left(i\omega_{n},\fvec k-\frac{\fvec q}{2}\right)\nonumber \\
 & \times G_{d}\left(i\omega_{n}+i\Omega_{m},\fvec k+\frac{\fvec q}{2}+\fvec Q\right)\nonumber \\
= & \frac{\lambda^{2}}{2|\fvec v_{c}\times\fvec v_{d}|}T\sum_{n}\frac{\mathrm{sgn}(\omega_{m}(\omega_{m}+\Omega_{n}))}{1+\frac{1}{2\Upsilon\tau}\mathrm{sgn}(\omega_{m}(\omega_{m}+\Omega_{n}))}
\end{align}

Therefore, the particle-hole bubble does not depend on $\fvec q$.
In the static limit, we find:
\begin{equation}
\Pi(0)=\frac{\lambda^{2}\Omega_{c}}{2\pi|\fvec v_{c}\times\fvec v_{d}|}\left(1+\frac{1}{2\Upsilon\tau}\right)^{-1}
\end{equation}

Thus, since the correlation length is given by $\xi^{-2}=r_{0}-\chi_{0}\Pi\left(0\right)$,
and using the definition $r=\frac{\xi^{-2}\gamma}{2\pi}$, we find:
\begin{equation}
\frac{\rmd r}{\rmd\tau^{-1}}=\frac{\Omega_{c}}{4\pi\Upsilon}\label{Eqn:r}
\end{equation}

The above result is consistent with previous works that found a reduction
of the magnetic order parameter with disorder in itinerant AFM systems
\cite{RMF12,Dagotto15}. We can also calculate the correction to the
Landau damping $\gamma^{-1}\equiv\chi_{0}\left[\Pi(0)-\Pi(i\Omega_{n})\right]/\left|\Omega_{n}\right|$.
We find:
\begin{align}
 & \Pi(0)-\Pi(i\Omega_{n})\nonumber \\
= & \frac{\lambda^{2}}{2|\fvec v_{c}\times\fvec v_{d}|}T\sum_{m=1}^{|\Omega_{n}|/(2\pi T)}\left[\left(1+\frac{1}{2\Upsilon\tau}\right)^{-1}+\left(1-\frac{1}{2\Upsilon\tau}\right)^{-1}\right]\nonumber \\
= & \frac{\lambda^{2}|\Omega_{n}|}{4\pi|\fvec v_{c}\times\fvec v_{d}|}\left[\left(1+\frac{1}{2\Upsilon\tau}\right)^{-1}+\left(1-\frac{1}{2\Upsilon\tau}\right)^{-1}\right]
\end{align}
yielding:
\begin{equation}
\gamma^{-1}=\frac{\lambda^{2}\chi_{0}}{2\pi|\fvec v_{c}\times\fvec v_{d}|}\left(1-(2\Upsilon\tau)^{-2}\right)^{-1}\ ,\quad\frac{\rmd\gamma^{-1}}{\rmd\tau^{-1}}=0
\end{equation}
Therefore, the Landau damping $\gamma$ depends only quadratically
on the scattering rate, and does not contribute to the leading order
in $\tau^{-1}$.

\subsection{Total suppression rate of $T_{c}$}

Using the results of the previous sections, we can rewrite the matrix
elements in Eqs.~(\ref{Eqn:KOffDia} and \ref{Eqn:KDia}) as:
\begin{align}
 & \hat{K}_{m\neq n}=\sqrt{\frac{\Lambda}{T}}\left(\frac{f_{m}f_{n}(\lambda_{+}/\lambda)^{2}}{\sqrt{|n-m|+r/T}}+\frac{f_{m}f_{n}(\lambda_{-}/\lambda)^{2}}{\sqrt{n+m+1+r/T}}\right)\nonumber \\
 & \hat{K}_{nn}=\frac{2(\lambda_{-}/\lambda)^{2}f_{n}^{2}}{\sqrt{2n+1+r/T}}\sqrt{\frac{\Lambda}{T}}-\pi(2n+1)-\frac{1}{\tau_{Q}T}-\nonumber \\
 & \quad f_{n}\sqrt{\frac{\Lambda}{T}}\sum_{\substack{m\neq n\\
m=0
}
}^{N_{c}}f_{m}\left[\frac{(\lambda_{+}/\lambda)^{2}}{\sqrt{|n-m|+r/T}}-\frac{(\lambda_{-}/\lambda)^{2}}{\sqrt{n+m+1+r/T}}\right]
\end{align}

While $\pd\eta/\pd T$ is the same as Eq.~(\ref{Eqn:TDerivative}),
the term $\pd\eta/\pd\tau^{-1}$ acquires two new contributions arising
from the vertex renormalization ($b_{1}$) and from the self-energy
renormalization ($b_{2}$):

\begin{equation}
\left(\frac{\pd\eta}{\pd\tau^{-1}}\right)_{b}=\left(\frac{\pd\eta}{\pd\tau^{-1}}\right)_{b,1}+\left(\frac{\pd\eta}{\pd\tau^{-1}}\right)_{b,2}
\end{equation}

Using Eq.~(\ref{Eqn:lambda}), we find the contribution from the
vertex renormalization: 
\begin{align}
 & \Upsilon\left(\frac{\pd\hat{K}_{m\neq n}}{\pd\tau^{-1}}\right)_{b,1}=\sqrt{\frac{\Lambda}{T}}f_{m}f_{n}\left[\frac{-1}{\sqrt{|n-m|+r/T}}\right.\nonumber \\
 & \quad\left.+\frac{1}{\sqrt{n+m+1+r/T}}\right]\nonumber \\
 & \Upsilon\left(\frac{\pd\hat{K}_{nn}}{\pd\tau^{-1}}\right)_{b,1}=2f_{n}^{2}\sqrt{\frac{\Lambda}{T}}\left(2n+1+r/T\right)^{-1/2}+\nonumber \\
 & \quad\sqrt{\frac{\Lambda}{T}}\sum_{\substack{m\neq n\\
m=0
}
}f_{m}f_{n}\left(\frac{1}{\sqrt{|n-m|+r/T}}+\frac{1}{\sqrt{n+m+1+r/T}}\right)\label{aux_b1}
\end{align}
Similarly, we find the contribution from the bosonic self-energy renormalization
\begin{align}
 & \Upsilon\left(\frac{\pd\hat{K}_{m\neq n}}{\pd\tau^{-1}}\right)_{b,2}=-f_{m}f_{n}\frac{\Omega_{c}}{8\pi T}\sqrt{\frac{\Lambda}{T}}\left[\left(|n-m|+r/T\right)^{-\frac{3}{2}}\right.\nonumber \\
 & \quad\left.+\left(n+m+1+r/T\right)^{-\frac{3}{2}}\right]\nonumber \\
 & \Upsilon\left(\frac{\pd\hat{K}_{nn}}{\pd\tau^{-1}}\right)_{b,2}=-2f_{n}^{2}\frac{\Omega_{c}}{8\pi T}\sqrt{\frac{\Lambda}{T}}\left(2n+1+r/T\right)^{-\frac{3}{2}}+\nonumber \\
 & \quad\frac{\Omega_{c}}{8\pi T}\sqrt{\frac{\Lambda}{T}}\sum_{\substack{m\neq n\\
m=0
}
}f_{m}f_{n}\left[\left(|n-m|+r/T\right)^{-\frac{3}{2}}-\right.\nonumber \\
 & \hspace{3cm}\left.\left(n+m+1+r/T\right)^{-\frac{3}{2}}\right]\label{aux_b2}
\end{align}

It is now straightforward to compute $\left(\frac{\mathrm{d}T_{c}}{\mathrm{d}\tau^{-1}}\right)_{b,1}$
and $\left(\frac{\mathrm{d}T_{c}}{\mathrm{d}\tau^{-1}}\right)_{b,2}$
numerically, using the solution of the clean system obtained in the
previous section. The red curve in Fig.~\ref{Fig:Boson}(a) shows
$\left(\frac{\mathrm{d}T_{c}}{\mathrm{d}\tau^{-1}}\right)_{b,1}$
as function of the distance to the QCP, whereas the blue curve shows
$\left(\frac{\mathrm{d}T_{c}}{\mathrm{d}\tau^{-1}}\right)_{b,2}$.
Surprisingly, not only the former is larger in magnitude than the
latter, but it is also positive, whereas the latter is negative. The
result $\left(\frac{\mathrm{d}T_{c}}{\mathrm{d}\tau^{-1}}\right)_{b,2}<0$
is straightforward to understand qualitatively: because $r\propto\xi^{-2}$
is enhanced by disorder, according to Eq. (\ref{Eqn:r}), the system
behaves as it moves away from the QCP, which effectively reduces $T_{c}$,
according to the behavior found previously in the clean system in
Fig.~\ref{Fig:QCPSC}. On the other hand, the result $\left(\frac{\mathrm{d}T_{c}}{\mathrm{d}\tau^{-1}}\right)_{b,2}>0$
is less straightforward to understand qualitatively, particularly
since disorder may enhance or suppress the vertex $\lambda$ depending
on the frequencies of the two external fermions, as shown by Eq. (\ref{Eqn:lambda}).

This unexpected result can be understood by analyzing the expression
for the coherent spectral weigh $Z_{n}^{-1}$, Eq.~(\ref{Eqn:EliashbergZ}),
in the presence of the renormalized electron-boson coupling $\lambda_{r}$
(and in the absence of other impurity terms). At the QCP, we find
that at low frequencies, $\omega\ll\Lambda_{c}$, $Z_{n}$ is effectively
reduced by this vertex renormalization, $\left(\frac{\rmd Z}{\rmd\tau^{-1}}\right)_{b_{1}}=-\frac{1}{\Upsilon|\omega|}\sqrt{\frac{\Lambda\Omega_{c}}{2\pi}}$.
Consequently, because $Z_{n}$ appears in the denominator of the pairing
kernel in the gap equation (\ref{Eqn:EliashbergGap}), the SC transition
temperature is enhanced by this effect. Note that the pairing kernel
also has a factor of $\lambda^{2}$ in the numerator; however, because
the sign of the vertex correction $\delta\lambda$ in Eq. (\ref{delta_lambda})
changes depending on the relative frequencies of the external fermions,
it does not compensate for the effect arising from the suppression
of $Z_{n}$ in the denominator. Indeed, the only reason $Z_{n}$ is
efficiently suppressed by $\lambda^{2}$ is because of the term $\mathrm{sign}\left(\omega_{m}\right)$
inside the sum in Eq.~(\ref{Eqn:EliashbergZ}), which is compensated
by the same term $\mathrm{sign}\left(\omega_{m}\right)$ in Eq.~(\ref{Eqn:lambdaSign}).
Such compensation leads to the cutoff dependence of $\rmd Z/\rmd\tau^{-1}$,
and outweighes the impact of disorder on the renormalized pairing
kernel.

Analytically, we can obtain approximate expressions for both $\left(\frac{\mathrm{d}T_{c}}{\mathrm{d}\tau^{-1}}\right)_{b,1}$
and $\left(\frac{\mathrm{d}T_{c}}{\mathrm{d}\tau^{-1}}\right)_{b,2}$
at the QCP, $r=0$. The details are shown in Appendix A, and give:

\begin{align}
\frac{\Upsilon}{\Lambda}\left(\frac{\dif T_{c}}{\dif\tau^{-1}}\right)_{b,1} & \approx0.6\sqrt{\frac{\Omega_{c}}{\Lambda}}\nonumber \\
\frac{\Upsilon}{\Lambda}\left(\frac{\dif T_{c}}{\dif\tau^{-1}}\right)_{b,2} & \approx-0.045\frac{\Omega_{c}}{\Lambda}\label{analytics}
\end{align}

The reason why $\left(\frac{\mathrm{d}T_{c}}{\mathrm{d}\tau^{-1}}\right)_{b,2}$
grows faster with $\Omega_{c}$ is because high-energy states contribute
more to the particle-hole bubble than to the vertex correction. In
Fig. \ref{Fig:Boson}(d) we present the net result $\frac{\Upsilon}{\Lambda}\left(\frac{\dif T_{c}}{\dif\tau^{-1}}\right)_{b,1}+\frac{\Upsilon}{\Lambda}\left(\frac{\dif T_{c}}{\dif\tau^{-1}}\right)_{b,2}$
as function of the distance to the QCP and of the cutoff. Clearly,
for a wide regime of parameters the net effect of the coupling between
disorder and bosonic degrees of freedom is an enhancement of $T_{c}$.

\section{Concluding Remarks}

In this work, we used a variational approach to investigate how different
effects contribute to the suppression rate of $T_{c}$ by disorder,
$\frac{\rmd T_{c}}{\rmd\tau^{-1}}$, in the case of an unconventional
superconductor in which pairing is mediated by quantum critical fluctuations.
By studying the spin-fermion model in the large-$N$ hot-spot approximmation,
we identified three different contributions to the reduction of $T_{c}$
with impurity scattering, $\frac{\rmd T_{c}}{\rmd\tau^{-1}}=\left(\frac{\rmd T_{c}}{\rmd\tau^{-1}}\right)_{f}+\left(\frac{\rmd T_{c}}{\rmd\tau^{-1}}\right)_{b,1}+\left(\frac{\rmd T_{c}}{\rmd\tau^{-1}}\right)_{b,2}$,
as outlined in Eq. (\ref{main}). $\left(\frac{\rmd T_{c}}{\rmd\tau^{-1}}\right)_{f}$
arises from the pair-breaking effect promoted by the coupling between
the electrons and the large-momentum impurity potential. As shown
in Fig.~\ref{Fig:TcInf}, $\left(\frac{\rmd T_{c}}{\rmd\tau^{-1}}\right)_{f}$
is always negative albeit reduced with respect to the Abrikosov-Gor'kov
value near the QCP. Such a reduction stems from the suppression of
quasi-particle spectral weight near the QCP, and has its roots on
the non-Fermi liquid character of the AFM QCP.

While this trend agrees with results from previous works on similar
spin-fermion models \cite{Norman93,AVC10}, our variational approach,
by means of the Hellmann-Feynman theorem, allows us to also assess
the effect of the coupling between disorder and the pairing interaction
(i.e. the bosonic degrees of freedom) without having to solve the
complicated disordered problem. Two contributions arise: $\left(\frac{\rmd T_{c}}{\rmd\tau^{-1}}\right)_{b,1}$,
due to the dressing of the electron-boson vertex by impurities, and
$\left(\frac{\rmd T_{c}}{\rmd\tau^{-1}}\right)_{b,2}$ due to the
dressing of the bosonic self-energy by impurities. Surprisingly, we
find $\left(\frac{\rmd T_{c}}{\rmd\tau^{-1}}\right)_{b,1}>0$ close
to the QCP and larger in magnitude than $\left(\frac{\rmd T_{c}}{\rmd\tau^{-1}}\right)_{b,2}<0$.
While the latter behavior can be understood as a result of the suppression
of the magnetic correlation length by disorder, the former stems from
the enhancement of the quasi-particle spectral weight promoted by
the renormalization of the electron-boson coupling.

It is interesting to discuss the relative magnitudes of these effects.
Our analytical approximations, combined with the numerical results,
show that at the QCP the two effects arising from the coupling of
disorder to the bosons behave as $\left|\frac{\rmd T_{c}}{\rmd\tau^{-1}}\right|_{b,2}\sim\frac{0.04}{\sin\theta}\left|\frac{\rmd T_{c}}{\rmd\tau^{-1}}\right|_{b,1}$.
Therefore, unless the system is very close to perfect nesting ($\theta=0$),
the positive contribution $\left(\frac{\rmd T_{c}}{\rmd\tau^{-1}}\right)_{b,1}$
overcomes the negative contribution $\left(\frac{\rmd T_{c}}{\rmd\tau^{-1}}\right)_{b,2}$,
as illustrated in Fig.~\ref{Fig:Boson}(d). Consequently, the suppression
rate of $T_{c}$ enforced by the direct coupling of the fermions to
the impurity potential $\left(\frac{\rmd T_{c}}{\rmd\tau^{-1}}\right)_{f}$
is even more reduced as compared to the Abrikosov-Gor'kov value. In
particular, we can estimate using our analytical expressions $\left|\frac{\rmd T_{c}}{\rmd\tau^{-1}}\right|_{b,1}\sim\frac{\lambda^{2}\chi_{0}}{E_{F}}\left|\frac{\rmd T_{c}}{\rmd\tau^{-1}}\right|_{f}$,
implying that this additional enhancement of $T_{c}$ is generally
smaller than the reduction promoted by pair-breaking effects. Equivalently,
within an expansion in the number of hot spots $N$, this additional
contribution acquires a prefactor of $1/\sqrt{N}$. Thus, the universal
value $\left(\frac{\rmd T_{c}}{\rmd\tau_{Q}^{-1}}\right)_{f}\approx-0.45$
obtained at the QCP (i.e. the value obtained when $r=0$ and $\Omega_{c}\rightarrow\infty$)
is an upper boundary value that may in principle be used to test this
model. Experimentally, it would be interesting to obtain $\left(\frac{\rmd T_{c}}{\rmd\tau^{-1}}\right)$
experimentally in electron-doped cuprates or iron pnictides near the
putative AFM QCP by introducing disorder in a controlled way via,
for instance, irradation.

In summary, we have shown that the suppression of $T_{c}$ by weak
disorder in an AFM quantum critical SC is significantly reduced compared
to the universal value obtained from the Abrikosov-Gor'kov theory
of conventional dirty SC. Our work highlights the importance of the
incoherent electronic spectral weight and of the feedback of the electronic
states on the pairing interaction to describe the properties of this
unconventional pairing state. Qualitatively, our results agree with
several experimental observations in cuprates and pnictides reporting
a robust SC state against disorder. Extensions of this promising framework
to include higher-order contributions from the impurity scattering
would be desirable to achieve more quantitative comparisons with experiments,
such as the critical value of the impurity scattering that destroys
the quantum critical SC state.
\begin{acknowledgments}
We thank A. Chubukov, A. Millis, J. Schmalian, X. Wang, and Y. Wang
for fruitful discussions. This work was supported by the U.S. Department
of Energy, Office of Science, Basic Energy Sciences, under award number
DE-SC0012336. \end{acknowledgments}

\appendix
\setcounter{figure}{0} \setcounter{table}{0} \global\long\def\theequation{S\arabic{equation}}
 \global\long\def\thefigure{S\arabic{figure}}

\section{analytical calculation of $\left(\frac{\mathrm{d}T_{c}}{\mathrm{d}\tau^{-1}}\right)_{b,1}$
and $\left(\frac{\mathrm{d}T_{c}}{\mathrm{d}\tau^{-1}}\right)_{b,2}$
at the QCP}

Here we focus on the case where the system is at the QCP, $r=0$,
and apply the hard cutoff procedure for $N_{c}\gg1$. We start by
computing $\left(\frac{\rmd T_{c}}{\rmd\tau^{-1}}\right)_{b,1}$.
In this case, the expressions (\ref{aux_b1}) simplify to:
\begin{align}
\Upsilon\left(\frac{\pd K_{nn}}{\pd\tau^{-1}}\right)_{b,1}= & \sqrt{\frac{\Lambda}{T}}\left[2\zeta\left(\half,1\right)-\zeta\left(\half,N_{c}-n\right)\right.\nonumber \\
 & \left.-\zeta\left(\half,N_{c}+n\right)+\frac{1}{\sqrt{2n+1}}\right]\nonumber \\
\Upsilon\left(\frac{\pd K_{m\neq n}}{\pd\tau^{-1}}\right)_{b,1}= & -\sqrt{\frac{\Lambda}{T}}\left[\frac{1}{\sqrt{|m-n|}}-\frac{1}{\sqrt{m+n+1}}\right]
\end{align}
where $\zeta\left(a,x\right)$ is the Hurwitz zeta function. In the
limit $N_{c}\gg1$, the off-diagonal term is much smaller than the
diagonal one. Consequently, the change in the eigenvalue is given
by:

\begin{equation}
\Upsilon\sqrt{\frac{T}{\Lambda}}\left(\frac{\dif\eta}{\dif\tau^{-1}}\right)_{b,1}=2\sum_{n=0}^{N_{c}}\bar{\Delta}_{n}^{2}\left(\sqrt{N_{c}-n}+\sqrt{N_{c}+n}\right)
\end{equation}
where the eigenvectors are normalized, $\sum_{n}\bar{\Delta}_{n}^{2}=1$.
As discussed in the solution of the clean case, $\bar{\Delta}_{n\geq n_{0}}\approx An^{-3/2}$
for $1\ll n_{0}\ll N_{c}$, with $A>0$. Thus, we obtain:
\begin{align}
\Upsilon\sqrt{\frac{T}{\Lambda}}\left(\frac{\dif\eta}{\dif\tau^{-1}}\right)_{b,1} & \approx4\sqrt{N_{c}}\sum_{n=0}^{n_{0}}\bar{\Delta}_{n}^{2}\nonumber \\
 & +2A^{2}\int_{n_{0}}^{N_{c}}dn\left(\frac{\sqrt{N_{c}-n}+\sqrt{N_{c}+n}}{n^{3}}\right)\nonumber \\
 & \approx4\sqrt{N_{c}}\sum_{n=0}^{n_{0}}\bar{\Delta}_{n}^{2}+\frac{2A^{2}\sqrt{N_{c}}}{n_{0}^{2}}
\end{align}
yielding:

\begin{equation}
\Upsilon\left(\frac{\dif\eta}{\dif\tau^{-1}}\right)_{b,1}\approx\frac{8}{\sqrt{2\pi}}\sqrt{\frac{\Omega_{c}}{\Lambda}}\ .\label{S_db1}
\end{equation}
where we used the clean limit result $T_{c}\approx\Lambda/2$. Therefore,
$T_{c}$ actually increases due to the dressing of the fermon-boson
vertex by disorder. To evaluate the change in $T_{c}$ due to this
effect, we use Eq. (\ref{Eqn:Delta_dT}):
\begin{equation}
-\left(\frac{\dif\eta}{\rmd T}\right)=-\frac{\pi}{2T}\sum_{n=0}^{\infty}\bar{\Delta}_{n}^{2}(2n+1)\approx\frac{1.6\pi}{\Lambda}\label{s_db1_aux}
\end{equation}
where the last step was obtained by the numerical solution of the
clean system at the QCP. Therefore, we obtain:

\begin{equation}
\frac{\Upsilon}{\Lambda}\left(\frac{\dif T_{c}}{\dif\tau^{-1}}\right)_{b,1}\approx0.6\sqrt{\frac{\Omega_{c}}{\Lambda}}\label{S_Tc_b1}
\end{equation}

\begin{figure}[htbp]
\subfigure[\label{FigS:QCP:DerTcCouple}]{\includegraphics[scale=0.7]{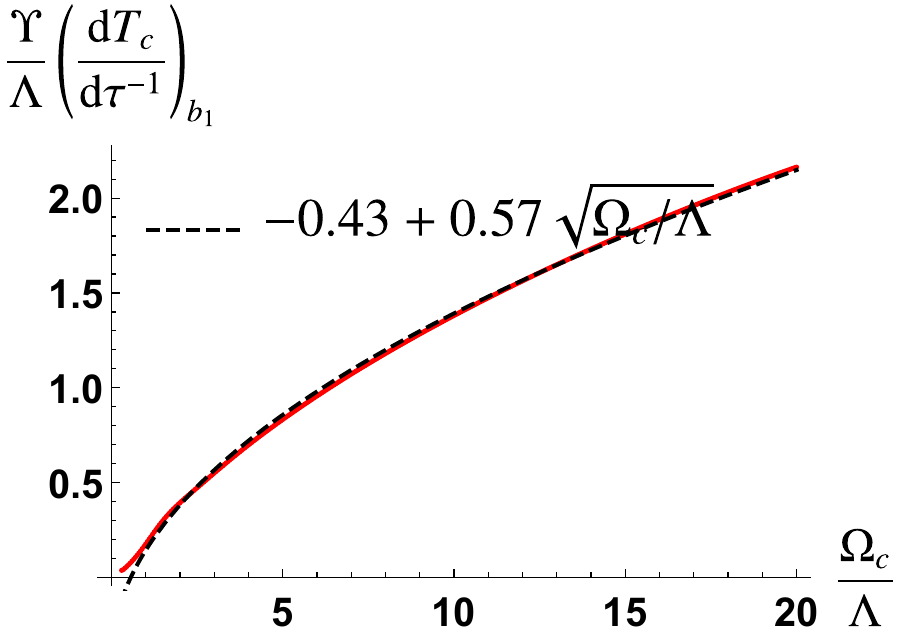}}
\hspace{1cm} \subfigure[\label{FigS:QCP:DerTcSE}]{\includegraphics[scale=0.7]{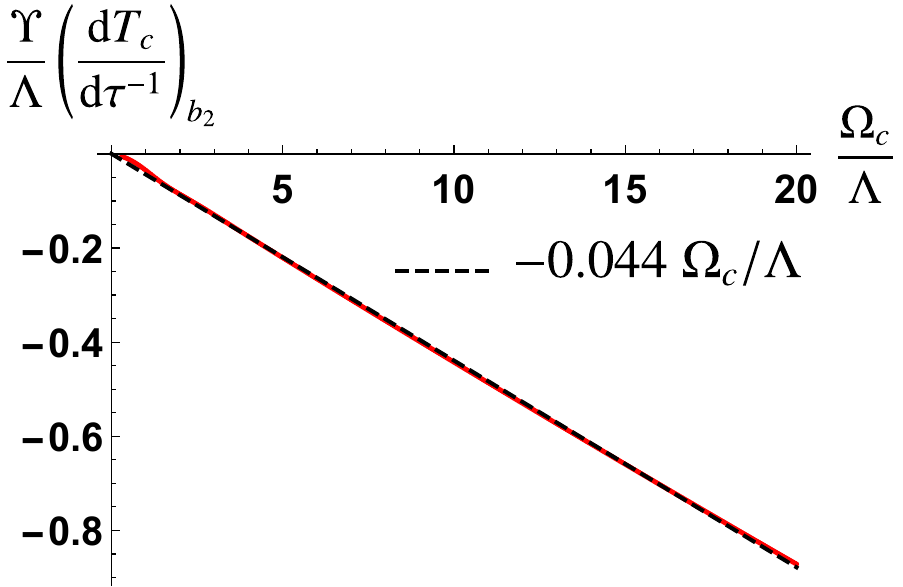}}
\protect\protect\protect\protect\protect\caption{Contribution to the suppression rate $\left(\dif T_{c}/\dif\tau^{-1}\right)$
arising from the impurity dressing of (a) the fermion-boson coupling
and (b) the bosonic self-energy. Solid curves are the numerical result,
dashed curves are the analytical approximations.}

\label{FigS:QCP}
\end{figure}

This approximate analytical expression is in very good agreement with
the numerical results, as shown in Fig.~\ref{FigS:QCP}(a).

We now move on to compute $\left(\frac{\rmd T_{c}}{\rmd\tau^{-1}}\right)_{b,2}$
at the QCP. From Eq. (\ref{aux_b2}) we have, for $r=0$:
\begin{align}
\Upsilon\left(\frac{\partial\hat{K}_{nn}}{\partial\tau^{-1}}\right)_{b,2}= & \frac{\Omega_{c}}{4\pi T}\sqrt{\frac{\Lambda}{T}}\left[\zeta\left(\frac{3}{2},1\right)-\zeta\left(\frac{3}{2},n+1\right)\right.\nonumber \\
 & \left.-\frac{1}{2(2n+1)^{3/2}}\right]\nonumber \\
\Upsilon\left(\frac{\partial\hat{K}_{m\neq n}}{\partial\tau^{-1}}\right)_{b,2}= & -\half\frac{\Omega_{c}}{4\pi T}\sqrt{\frac{\Lambda}{T}}\left[\frac{1}{|m-n|^{3/2}}+\right.\nonumber \\
 & \left.\frac{1}{(m+n+1)^{3/2}}\right]
\end{align}

As a result, using the same procedure as above, we obtain:
\begin{align}
 & \Upsilon\sqrt{\frac{T}{\Lambda}}\frac{4\pi T}{\Omega_{c}}\left(\frac{\dif\eta}{\dif\tau^{-1}}\right)_{b,2}=\sum_{n}\bar{\Delta}_{n}^{2}\left[\zeta\left(\frac{3}{2},1\right)-\zeta\left(\frac{3}{2},n+1\right)\right]\nonumber \\
 & \qquad-\half\sum_{m,n}\bar{\Delta}_{m}\bar{\Delta}_{n}\left[\frac{1-\delta_{mn}}{|m-n|^{3/2}}+\frac{1}{(m+n+1)^{3/2}}\right]
\end{align}

Using the numerical solution of the clean system at the QCP, we find
for the right-hand side of the equation:

\begin{equation}
\Upsilon\sqrt{\frac{T}{\Lambda}}\frac{4\pi T}{\Omega_{c}}\left(\frac{\dif\eta}{\dif\tau^{-1}}\right)_{b,2}\approx-1
\end{equation}
yielding:

\begin{equation}
\frac{\Upsilon}{\Lambda}\left(\frac{\dif T_{c}}{\dif\tau^{-1}}\right)_{b,2}\approx-0.045\frac{\Omega_{c}}{\Lambda}\label{S_Tc_b2}
\end{equation}

As shown in Fig.~\ref{FigS:QCP}(b), the numerical results agree
well with this expression.
\end{document}